# Control of microswimmers by spiral nematic vortices: transition from individual to collective motion and contraction, expansion, and stable circulation of bacterial swirls

Runa Koizumi[1], Taras Turiv[1], Mikhail M. Genkin[2], Robert J. Lastowski[1,3], Hao Yu[1], Irakli Chaganava[4,5], Qi-Huo Wei[1,6], Igor S. Aranson[7], Oleg D. Lavrentovich[1,6,*]

[1] *Advanced Materials and Liquid Crystal Institute, Materials Science Graduate Program, Kent State University, Kent, OH 44242, USA*
[2] *Cold Spring Harbor Laboratory, 1 Bungtown Rd, Cold Spring Harbor, NY 11724, USA*
[3] *Department of Chemistry, University of Scranton, Scranton, PA, 18510, USA*
[4] *Institute of Cybernetics of Georgian Technical University, Tbilisi 0186, Georgia*
[5] *Georgian State Teaching University of Physical Education and Sport, Tbilisi 0162, Georgia*
[6] *Department of Physics, Kent State University, Kent, OH 44242, USA*
[7] *Department of Biomedical Engineering, Pennsylvania State University, University Park, PA 16802, USA*

* Author for correspondence: Oleg D. Lavrentovich

**Email:** *olavrent@kent.edu*

**Abstract**

Active systems comprised of self-propelled units show fascinating transitions from Brownian-like dynamics to collective coherent motion. Swirling of swimming bacteria is a spectacular example. This study demonstrates that a nematic liquid crystal environment patterned as a spiral vortex controls individual-to-collective transition in bacterial swirls and defines whether they expand or shrink. In dilute dispersions, the bacteria swim along open spiral trajectories, following the pre-imposed molecular orientation. The trajectories are nonpolar. As their concentration exceeds some threshold, the bacteria condense into unipolar circular swirls resembling stable limit cycles. This collective circular motion is controlled by the spiral angle that defines the splay-to-bend ratio of the background director. Vortices with dominating splay shrink the swirls towards the center, while vortices with dominating bend expand them to the periphery. $45^o$ spiraling vortices with splay-bend parity produce the most stable swirls. All the dynamic scenarios are explained by hydrodynamic interactions of bacteria mediated by the patterned passive nematic environment and by the coupling between the concentration and orientation. The acquired knowledge of how to control individual and collective motion of microswimmers by a nematic environment can help in the development of microscopic mechanical systems.

## I. INTRODUCTION

Aqueous dispersions of self-propelled microswimmers such as motile flagellated bacteria that convert stored energy into mechanical motion are one of the most studied examples of active matter [1]. Recent reports demonstrate intriguing examples of collective effects at length scales much larger than the size of an individual swimmer, such as coordinated motion and self-concentration caused by hydrodynamic and steric interactions of bacteria among themselves and with bounding surfaces [2-7]. These studies are of importance not only for the fundamental understanding of active matter [1,8] but also for potential future applications, since the power of collective motion can be harnessed to drive micromachines [9-11]. To realize the full potential of the collective motion, one needs to learn how to control it. In an aqueous environment, a natural habitat of most microswimmers, this control can be achieved either by the gradients of nutrients which are usually transient and thus do not allow to support a sustained motion or by physical confinement such as microchambers, solid inclusions or obstacles [3,9,10,12-14]. Another solution is to replace the isotropic aqueous medium with an anisotropic liquid crystalline environment that can guide the individual [15-17] and collective [17,18] dynamics of bacteria [19]. Bacterial dispersions in lyotropic chromonic liquid crystals (LCLCs), called "living liquid crystals" or "living nematics," exhibit an intriguing interplay of bacterial activity with the underlying orientational order of the passive nonpolar nematic environment, specified by the director $\hat{\mathbf{n}}$ with the properties $\hat{\mathbf{n}} \equiv -\hat{\mathbf{n}}$, $\hat{\mathbf{n}}^2 = 1$. In a uniformly aligned nematic, the rod-like bacteria such as *P. mirabilis* [20] and *B. subtilis* [17] prefer to swim along the director, as dictated by the elastic, surface anchoring and viscous interactions of their rod-like bodies with the surrounding anisotropic fluid [16-18,21]. When the nematic director is predesigned to be spatially-nonuniform [17,22-24], it can trigger unidirectional flows of bacteria and cause their strongly non-uniform spatial distribution [25]. In particular, Peng et al [23] demonstrated that $45°$ spiral vortices attract swimming bacteria and force them to swim along circular trajectories with a well-defined polarity. In contrast, circular vortices cause a qualitatively different bipolar circulation, with bacteria swimming both counterclockwise and clockwise in the same swirl [23]. This dramatic difference, coupled with the fact that circular swirling is one of the most common examples of coherent spatiotemporal behavior of bacteria [2,3,12,14,26,27] and other active units [28-30] calls for a systematic study.

In this work, we explore the onset and time development of collective swirling motion by imposing vortex director patterns of a background LCLC on the motile bacteria *B. subtilis*. *B. subtilis* is a rod-shaped microorganism that could swim in a nematic LCLC at $\sim 15 \mu m/s$ by coordinated rotation of about 10-20 helical filaments, called flagella; the head of the bacterium rotates in an opposite direction with a smaller frequency [17,31]. *B. subtilis* swimming in a nematic LCLC generates a "hydrodynamic force dipole" pattern [17,31], typical for many flagellated bacteria swimming in water [32]: the surrounding fluid is pushed away from the bacterium along the axial direction and pulled towards the bacterium along the two perpendicular directions, as schematically shown in Fig.1a. In dense isotropic aqueous dispersions, with concentration in excess of $10^{15}$ m$^{-3}$, these hydrodynamic dipolar perturbations lead to spectacular collective effects of "bacterial turbulence", with vortices and jets [2,33]. Theoretical modeling of these dense dispersions [34,35] reveals that the overlapping hydrodynamic force dipoles destabilize the mutual alignment of neighboring bacteria, making a parallel swimming motion unstable. Our work demonstrates that by replacing the isotropic environment with nematic vortices, which mediates the interactions of the hydrodynamic force dipoles through the underlying director, one can achieve a stable parallel polar swimming of bacteria with controllable scenarios, such as swirl expansion and swirl contraction.

The predesigned vortices range from radial to circular, with the spiral angle $\phi_0$ varying from 0 to 90°. We demonstrate that any spiral director vortex with $0 < \phi_0 < 90°$ congregates bacteria into a unidirectionally circulating swirl when the concentration of bacteria exceeds some threshold, $(10^{13}$-$10^{14})$ m$^{-3}$, which is 10-100 times smaller than the critical concentration at which "bacterial turbulence" sets in isotropic systems. This concentration threshold corresponds to the expected range of bacterial interactions in the nematic LCLC, at which the fluid perturbations around the neighboring swimmers start to overlap. Below the threshold, the bacteria show only an individual behavior, as they swim parallel to the director and do not condense into circular swirls. The spiral angle destines the long-term fate of the collective circular swimming above the threshold. In patterns with $\phi_0 < 45°$, as time goes by, the swirls shrink towards the center, while in $\phi_0 > 45°$ patterns, they expand. Vortices with $\phi_0 = 45°$ produce the most stable unidirectional circular motion of bacteria. The corresponding swirls are capable of self-healing: when the number of bacteria increases above some threshold and the circular trajectories start to undulate, the swirl releases the excess of bacteria to resume steady circulation. The experimental data are qualitatively explained by demonstrating that the hydrodynamic interactions among the bacteria and the ensuing active forces depend on the underlying director field. These interactions are described by an agent-based modeling and by an advection-diffusion model of a living liquid crystal [36] that incorporates the active forces created by the microswimmers.

## II. MATERIALS AND METHODS

### 1. Sample preparation.

We use *Bacillus subtilis* (wild-type strain 1085), which is a rod-shaped bacterium with a body length of 5-7 μm. The bacteria are initially grown on a lysogeny broth (LB, purchased from Teknova) agar plate, then transferred to a tube containing 10 mL of liquid medium terrific broth (TB, purchased from Sigma Aldrich). The tube is placed in a shaking incubator at 35°C and extracted after 7-9 hours, once the maximum saturation of bacterial concentration is reached. The saturation of the bacterial concentration is measured using a custom-made optical density meter. 0.05 mL of the liquid medium containing the bacteria is then extracted from the tube, centrifuged for 2 minutes at 10,000 rpm. After centrifugation, the liquid medium which does not contain bacteria is removed, and a 0.5 mL 14wt% solution of disodium cromoglycate (DSCG, purchased from Alfa Aesar) in TB is added to the bacteria and carefully pipetted up and down several times. TB provides nutrients to support motility of the bacteria, keeping the bacteria motile in a sealed cell with no additional oxygen supply for approximately 1-2 hours.

2. **Director patterning and experimental design.**

The bacterial solutions are confined in flat cells, formed by 1mm thick soda-lime glass sheets cleaned thoroughly in an ultrasonic bath for 30 minutes at 60°C, rinsed with isopropanol, then placed in an oven to dry for 10 minutes at 80°C. The glass is UV-ozone cured in a chamber for 10 minutes, then spin-coated with a solution of an azo-dye in dimethylformamide. We use either 0.5 wt% filtered solution of Brilliant Yellow dye (purchased from Sigma Aldrich) or SD-1 dye in dimethylformamide (purchased from Sigma Aldrich), depending on the humidity [37]. The glass plates are hard-baked in the oven for 30 minutes at 100°C to fully evaporate the solvent.

Photopatterning is used to impose the desired anchoring conditions onto the background LCLCs. The glass substrates with photoalignment dye layers are illuminated with a metal-halide X-Cite 120 lamp. The light beam is passing through a plasmonic photomask with nano-slits that produce a local polarization of light perpendicular to the long axes of slits. Upon irradiation, the spin-coated photosensitive dye molecules align perpendicularly to local polarization. The photoaligned dye coatings align the LCLC director; the director pattern reproduces the pattern on nano-slits in the plasmonic photomask. The glass substrates are carefully blown with a compressed nitrogen air gun and spin-coated with a toluene solution of 6.7wt% reactive mesogen RM-257 (purchased from Wilshire Tech.), containing also 0.35wt% photoinitiator Irgacure-651 (purchased from Ciba). Subsequently, the substrates are illuminated with a UV-lamp (365nm, 6W) for 30 minutes for polymerization of RM-257 to form a liquid crystal elastomer coating with the director that follows alignment of the azodye [38]. The patterned substrates are assembled into cells, being separated by spherical and cylindrical spacers of diameter 20 μm or 5 μm. The thickness of the cells is measured using a spectrometer. The LCLC containing the bacteria is injected into the cell using capillary forces, and the open sides are sealed using an epoxy to prevent evaporation. The sample is heated to 43°C using a hot stage (Linkam model PE94) to melt the LCLC into the isotropic phase, after which it was slowly cooled down to the nematic phase in order to achieve better alignment of the LCLCs. When the sample is slowly cooled from the isotropic phase, the nematic phase first forms as sublayers at the top and bottom plates, which merge to form a homogenous nematic phase. While the sample is in the biphasic state, the bacteria swim along the imposed nematic director near the two bounding surfaces and continue to swim along the director after complete transition to the nematic phase. The interfaces that are parallel to the plates do not swipe away bacteria. Occasionally, there might be a nematic-isotropic interface that is not parallel to the plates. Such an interface could swipe few bacteria from the field of view. However, the bacteria remain in the pool of the material and once the transition to the nematic is complete, they are free to swim anywhere, as dictated by the processes described in the paper. We verified and confirmed that the initial concentration and distribution of bacteria before and after the phase transition remain the same. The desired alignment of DSCG establishes itself 15-20 seconds after the sample transitions completely into the nematic phase. Therefore, in order to assure that there is full alignment, we waited 60 seconds after the phase transition before starting to acquire any data. Note here that the bacteria show a steady level of activity for only a limited time, about 30 min, after being injected into the cells; we thus could not extend a reliable analysis to time intervals much longer than 10 min.

The alignment patterns are identical at the top and bottom surfaces. The pre-imposed director writes in cylindrical coordinates as

$$\hat{\mathbf{n}}_0 = (n_r, n_\phi, 0) = (\cos\phi_0, \sin\phi_0, 0), \qquad (1)$$

where the constant spiral angle $\phi_0$ is measured between the radius-vector $\mathbf{r}$ and the local director. Nine different patterns were designed, with $\phi_0 = 0$, $10°, 25°, 45°, 60°, 70°, 75°, 80°$ and $90°$, Figs. 1,2. The size of the square patterned area is 1 mm². The thickness $d$ of the cells is $20\,\mu\text{m}$, unless specified otherwise. Since $d$ is much smaller than the in-plane scale of deformations, the system is effectively two-dimensional. When $\phi_0 = 0$, the director deformation is a pure splay, $\mathbf{s} = \hat{\mathbf{n}}_0 \nabla \cdot \hat{\mathbf{n}}_0 = (1/r, 0, 0)$. As $\phi_0$ increases above 0, the splay $\mathbf{s} = (\cos^2\phi_0/r, \sin 2\phi_0/2r, 0)$ is accompanied by bend, $\mathbf{b} = \hat{\mathbf{n}}_0 \times \nabla \times \hat{\mathbf{n}}_0 = (\sin^2\phi_0/r, -\sin 2\phi_0/2r, 0)$, until the deformation becomes a pure bend at $\phi_0 = 90°$, $\mathbf{b} = (1/r, 0, 0)$. When $\phi_0 = 45°$, the contributions of splay and bend are equal to each other, $|\mathbf{s}| = |\mathbf{b}| = 1/\sqrt{2}r$.

Besides the director pattern, another important factor controlling the dynamics of bacteria is their concentration. In addition to the overall concentration $c_v$ measured as a number of bacteria per unit volume, which does not depend on the particular director pattern, we introduce a local vortex-specific volume concentration $c_p$ measured by counting the bacteria within a box $\Delta x \times \Delta y \times d$ centered at the vortex core, where $\Delta x = \Delta y = 0.4\,\text{mm}$. We also calculate the volume fractions $\Phi_v$ and $\Phi_p$ corresponding to $c_v$ and $c_p$. We treat a bacterium as a cylindrical rod of a length $7\,\mu\text{m}$ and a diameter $0.7\,\mu\text{m}$ [17], so that its volume is approximately $v_B \approx 3\,\mu\text{m}^3$. $\Phi_v$ and $\Phi_p$ are calculated as the total volume of bacteria within the region of interest divided by the volume of that region. For selected patterns, in which a number $n_B$ of bacteria form a condensed toroidal swirl of an outer $r_{out}$ and inner $r_{in}$ radii and of the thickness $d$, we calculate the volume fraction of a condensate as $\Phi_{cond} = n_B v_B / \left[ \pi d \left( r_{out}^2 - r_{in}^2 \right) \right]$. The boundaries of swirls with the outer $r_{out}$ and inner $r_{in}$ radii are marked by red circles in Fig. 2, 4, and 6. The $\Phi_v, \Phi_p, \Phi_{cond}$ quantities are less precise as compared to $c_v$ and $c_p$ but help to understand the separation distances between the bacteria.

### 3. Optical Microscopy and PolScope measurements.

For data acquisition and recording, we use a Nikon TE-2000 inverted microscope and a high-resolution camera (Emergent HR-20000C). The bacteria appeared as higher-intensity objects on a lower background intensity of the transparent liquid-crystal field. We found the average image over all frames and subtracted it from each frame of the video. With this, the background liquid crystal intensity becomes even more suppressed and we segregated the individual bacteria using an intensity thresholding technique. In order to obtain the optical retardation map and underlying director field, the samples are observed using an optical polarizing microscope (Nikon E600) mounted with a Cambridge Research Incorporated Abrio LC-PolScope package. The PolScope uses 546 nm monochromatic illumination to map out the optical retardance $\Gamma$ and orientation of the slow axis of the sample, from which a 2D director field is reconstructed [39,40]. The maximum optical retardance that can be measured using the PolScope is 273 nm, but measurements above 240 nm do not yield accurate results [41]. Therefore, we restricted our retardance measurements to be range 0-240 nm by making thin cells of thickness $d = 5\,\mu\text{m}$.

To visualize the bacterial dynamics, we use bright-field optical microscopy in which the LCLC and swimming bacteria show different levels of transparency. The background transmitted light intensity is subtracted from the entire frame to make the LCLC environment "black", while the light intensity transmitted through the bacterial bodies is enhanced to the maximum "bright" level. The direction of bacterial motion is deduced by comparing subsequent video frames.

4. **Description of computational model.**

We validated our findings using the advection-diffusion model [36]. In this model, the nematic dynamics is described by the following equations:

$$(\partial_t + \mathbf{v} \cdot \nabla)\mathbf{Q} - \mathbf{S} - \Gamma \mathbf{H} + \mathbf{F}_{ext} = 0, \quad (2)$$

$$\nabla \cdot (\boldsymbol{\sigma}_s + \boldsymbol{\sigma}_a + \boldsymbol{\sigma}_{visc} + \boldsymbol{\sigma}_{act} - p\mathbf{I}) - \zeta \mathbf{v} = 0, \quad (3)$$

$$\partial_t c^{\pm} + \nabla \cdot \left( \pm V_0 \hat{\mathbf{n}} c^{\pm} + \mathbf{v} c^{\pm} \right) = \mp \frac{c^+ - c^-}{\tau} + D_c \nabla^2 c^{\pm}. \quad (4)$$

Here $V_0$ is the bacterium swimming speed, $D_c$ is the diffusion coefficient, and $\tau$ is the direction reversal time (if a bacterium reverses a direction, it leaves the population $c^+$ and becomes $c^-$ or vice versa). Our experiments show that the reversal time is typically quite large, of the order of 30−60 sec. In Eqs. (2) and (3), $\hat{\mathbf{n}}$ can differ from the prescribed $\hat{\mathbf{n}}_0$ because of the torques and forces imposed by the bacteria.

Equation (2) describes the evolution of the tensorial nematic order parameter $\mathbf{Q}$, where $\mathbf{v}$ is the fluid velocity, tensor $\mathbf{S}$ accounts for the nematic alignment with the fluid flow, $\mathbf{H}$ is the molecular field that accounts for the nematic elasticity. Equation (3) is the linear momentum balance, where $\boldsymbol{\sigma}_s$ and $\boldsymbol{\sigma}_a$ are the symmetric and antisymmetric elastic contributions, respectively, $\boldsymbol{\sigma}_{visc}$ is the viscous contribution, $\boldsymbol{\sigma}_{act} = -\lambda \mathbf{Q} c$ is the active stress that depends on bacterial concentration $c = c^+ + c^-$. The active stress generates realigning force $\mathbf{F}_{act} = \nabla \cdot \boldsymbol{\sigma}_{act}$. In turn, $\mathbf{F}_{ext}$ models nematic relaxation towards the anchoring direction $\hat{\mathbf{n}} - \frac{2a\mathbf{F}_{ext}}{\gamma_s}$. Here $p$ is fluid pressure, and the term $-\zeta \mathbf{v}$ is a viscous friction term. For additional details, see Ref. [25]. Equations (2) and (3) are complemented by two evolution equations for bacterial concentrations $c = c^+ + c^-$, Eq. (4). Each of the bacterial populations $c^\pm$ swims in opposite directions parallel to the director. Given the nematic angle $\phi$, there is freedom in choosing a $c^+$ orientation angle $\theta^+$ ($\theta^+ = \phi$ or $\theta^+ = \phi + \pi$). In our previous work [25], we defined $c^+$ to swim to the right (and correspondingly $c^-$ to the left) and relabeled concentrations in the proximity of vertical nematic orientations. In this work, we take advantage of spiral nematic patterns and define $c^+$ to always swim clockwise and $c^-$ – counterclockwise, which allows us to avoid non-physical bacterial clustering without $c^\pm$ relabeling. In this case, $c^-$ bacteria swim away from the pattern center, while $c^+$ bacteria swim towards the center.

### III. RESULTS

#### 1. *Individual-to-collective transition in swimming patterns.*

In dilute dispersions, $c_V \approx 0.8 \times 10^{12} \, \text{m}^{-3}$ (volume fraction $\Phi_v \approx 2.4 \times 10^{-6}$), the bacteria simply swim along $\hat{\mathbf{n}}_0$, as illustrated by six trajectories in the vortex pattern with $\phi_0 = 25°$ in Fig. 1b and Supplemental Movie S1 [42]. The bacteria in this dilute regime do not interact with each other and do not distort the surface-imposed director, similarly to the previous studies [17,18,20,21]. When the number of bacteria increases above ten, the trajectories start to change from spiral to circular, Fig. 1c and Supplemental Movie S2 [42]. The circular trajectories form a large angle

~ 65° with the director field $\hat{\mathbf{n}}_0$ pre-inscribed at the bounding plates, a behavior clearly at odds with steady swimming parallel to $\hat{\mathbf{n}}_0$ at low concentrations.

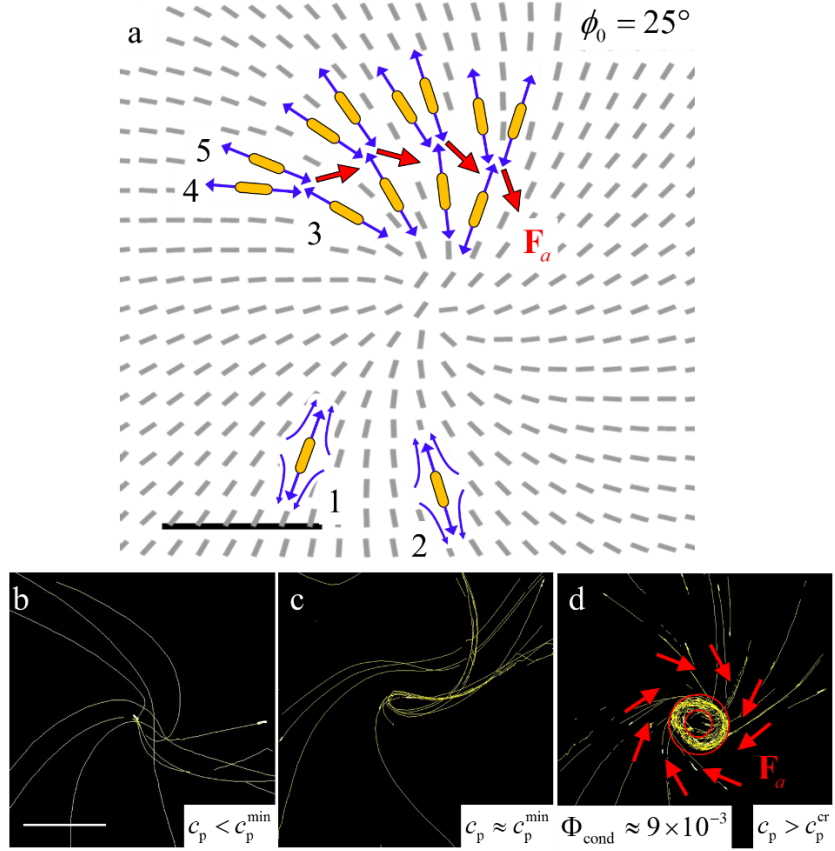

**Figure 1.** Transition from individual to collective motion caused by the increased bacterial concentration in a cell with a spiral vortex, $\phi_0 = 25°$. (a) Background: experimental director field $\hat{\mathbf{n}}_0$ mapped by PolScope in a cell with dilute dispersion ($c_v \approx 0.8 \times 10^{12} \, \text{m}^{-3}$, $\Phi_v \approx 2.4 \times 10^{-6}$). The overlaid scheme illustrates the transition from individual swimming (bacteria 1,2) to collective clockwise swirling (bacteria 3,4,5 and others nearby). Yellow rods represent swimming bacteria aligned by $\hat{\mathbf{n}}_0$. Blue arrows depict bacterial hydrodynamic force dipoles. Bacteria 1 and 2 do not interact, while bacteria 3,4,5 interact hydrodynamically. Overlapping of their force dipoles produces a net active force $\mathbf{F}_a$ shown by red arrows, which triggers clockwise circular swirling. (b) An optical microscopy texture with six overlapped individual trajectories that follow the surface-imposed $\hat{\mathbf{n}}_0$ at low concentration $c_p \approx 0.1 \times 10^{13} \, \text{m}^{-3}$, $\Phi_p \approx 3 \times 10^{-6}$. (c) Onset of collective

motion when the number of bacteria in the frame increases to 9: portions of trajectories become circular, $c_p^{min} \approx 0.2 \times 10^{13}$ m$^{-3}$, $\Phi_p \approx 6 \times 10^{-6}$. (d) A complete circulating swirl with $\approx 60$ bacteria collectively swimming clockwise at higher concentration $c_p \approx 0.6 \times 10^{14}$ m$^{-3}$, $\Phi_p \approx 1.8 \times 10^{-4}$, $\Phi_{cond} \approx 9 \times 10^{-3}$. Scale bar $50$ μm. The trajectories are tracked in the interval [0, 60 s].

The local concentration at the onset of the individual-to-collective motion transition, when only some but not all bacteria form circular or arched trajectories, Fig. 1c, is $c_p^{min} \approx 0.2 \times 10^{13}$ m$^{-3}$, $\Phi_p \approx 0.6 \times 10^{-5}$. The intermediate spiral-circular state persists until the concentration reaches a critical value $c_p^{cr} \approx 0.7 \times 10^{13}$ m$^{-3}$ (25-30 bacteria, $\Phi_p^{cr} \approx 2.1 \times 10^{-5}$), above which all bacteria condense into the circular swirl, as shown in Fig. 1d and Supplemental Movie S3 [42] for a group of ~60 bacteria, $c_p \approx 0.6 \times 10^{14}$ m$^{-3}$ and volume fraction $\Phi_{cond} \approx 9 \times 10^{-3}$ of bacteria in the condensed swirl. Importantly, in contrast to the individual regime that produces no net flows, the collective circulation is unipolar, namely, clockwise when the sample is viewed "from above", Fig. 1d. This unipolar circulation settles in all patterns with $0 < \phi_0 < 90°$, in which the director spirals counterclockwise from the center, as $\phi_0 > 0$ in Eq. (1), Fig. 1a. Occasionally, there might be a reversal of an individual bacterial velocity, but these events are rare.

The observed unipolar circulation is different from the well-known effect of circular swimming that many bacteria, including *B. subtilis*, experience in diluted dispersions, while swimming as individuals near solid boundaries: this individual circling is caused by the interactions of the bacterium's clockwise-rotating head with the substrate [43-45]. In our cells, the clockwise swimming is driven by the collective interactions between the bacteria mediated by the underlying spiraling director, which overcome any possible effects caused by the head counterrotation.

When the bacteria swim individually along $\hat{\mathbf{n}}_0$ in diluted samples, their average separation is large, $\approx 60\,\mu m$ or more. Collective circular swimming, Fig. 1c,d, occurs when the separation decreases to $\approx 15-20\,\mu m$, which corresponds to a significant overlap of the hydrodynamic force dipoles of neighboring bacteria in the LCLC background [31]. Hydrodynamic interactions of bacteria in proximity to each other, mediated by the patterned director field of the passive nematic environment, produce a net active force $\mathbf{F}_a$ that first triggers and then drives the collective motion, as illustrated in Fig. 1a and explained below.

Because no net force is applied to a self-propelled object, a swimming bacterium represents a moving hydrodynamic force dipole as it produces two outward fluid streams coaxial with the bacterial body [3,5,17], as shown for bacteria 1 and 2 in Fig.1a. The individual bacteria 1 and 2 are separated by large distances, their streams do not overlap, and they do not interact. These well separated bacteria experience only an alignment action of the passive nematic that orients them parallel to $\hat{\mathbf{n}}_0$ [15-18]. This alignment and nonpolar character of $\hat{\mathbf{n}}_0$ explain the spiral open nonpolar trajectories that the individual swimmers follow in Fig.1b.

The situation changes dramatically when the bacterial concentration increases and their force dipoles start to overlap, as illustrated in the top part of Fig.1a for bacteria 3,4,5 and others nearby. Since each bacterium prefers to be parallel to $\hat{\mathbf{n}}_0$, and since $\hat{\mathbf{n}}_0$ is predesigned with splay and bend, the force dipoles are titled with respect to each. The tilted force dipoles of closely located bacteria cannot compensate each other and produce a net active force. In the concrete example in Fig.1a with the spiral angle $\phi_0 = 25°$, bend and splay imposed onto the field of force dipoles, produce a net active force $\mathbf{F}_a$ that is directed clockwise and towards the center of the vortex, as shown by red arrows in Fig.1a,d. This net force triggers the transition from the individual to the collective clockwise circulation of the bacterial swarm. As we will see later, the qualitative picture of the bacterial interactions through the patterned director in Fig.1a,d is validated by a rigorous consideration of the active force dependence on the director gradients in Eq. (5), and by a theory that accounts for variable concentration.

The hydrodynamic interactions of overlapping dipoles in Fig.1a, mediated by the underlying passive director, also explain the dynamic phenomena that emerge when the transition to collective motion is complete. Among these are (i) clockwise circulation for any spiral angle in the range $0 < \phi_0 < 90°$; (ii) expansion and shrinkage of the swirls controlled by the spiral angle $\phi_0$. We described these effects below for samples with a higher concentration $c_v \approx 0.8 \times 10^{13}$ m$^{-3}$, $\Phi_v \approx 2.4 \times 10^{-5}$, Fig.2-4.

## 2. *Regime of steady unipolar circulation.*

All vortices with $0 < \phi_0 < 90°$ impose the same unidirectional clockwise steady circulation of bacteria, provided their concentration exceeds some critical value which depends on $\phi_0$ and is in the range $0.1 \times 10^{14}$ m$^{-3} \leq c_p \leq 0.8 \times 10^{14}$ m$^{-3}$ ($3 \times 10^{-5} \leq \Phi_p \leq 2.4 \times 10^{-4}$). The steady circulation persists till the concentration reaches a higher value $c_p^{max}$ at which the swirls start to undulate, losing their circular shape; this effect is detailed in the next section. The trajectories of bacteria tracked using TrackMate [46] are shown in Fig. 2a-q, using different colors for different bacteria.

The bacteria circulate within a relatively narrow ring, Fig. 2s (Supplemental Movie S3, S4 [42]). The radial distribution of bacteria in each pattern is tracked for 60 s in order to find the correlation between the radius of maximum concentration $r_{c,max}$ and $\phi_0$, which defines the ratio of bend to splay, Fig. 2s. The data from each pattern are normalized by setting the value of maximum concentration to 1. Each concentration distribution shows a prominent peak at $r = r_{c,max}$, Fig. 2s; $r_{c,max}$ increases with $\phi_0$, as the bend deformation overcomes splay.

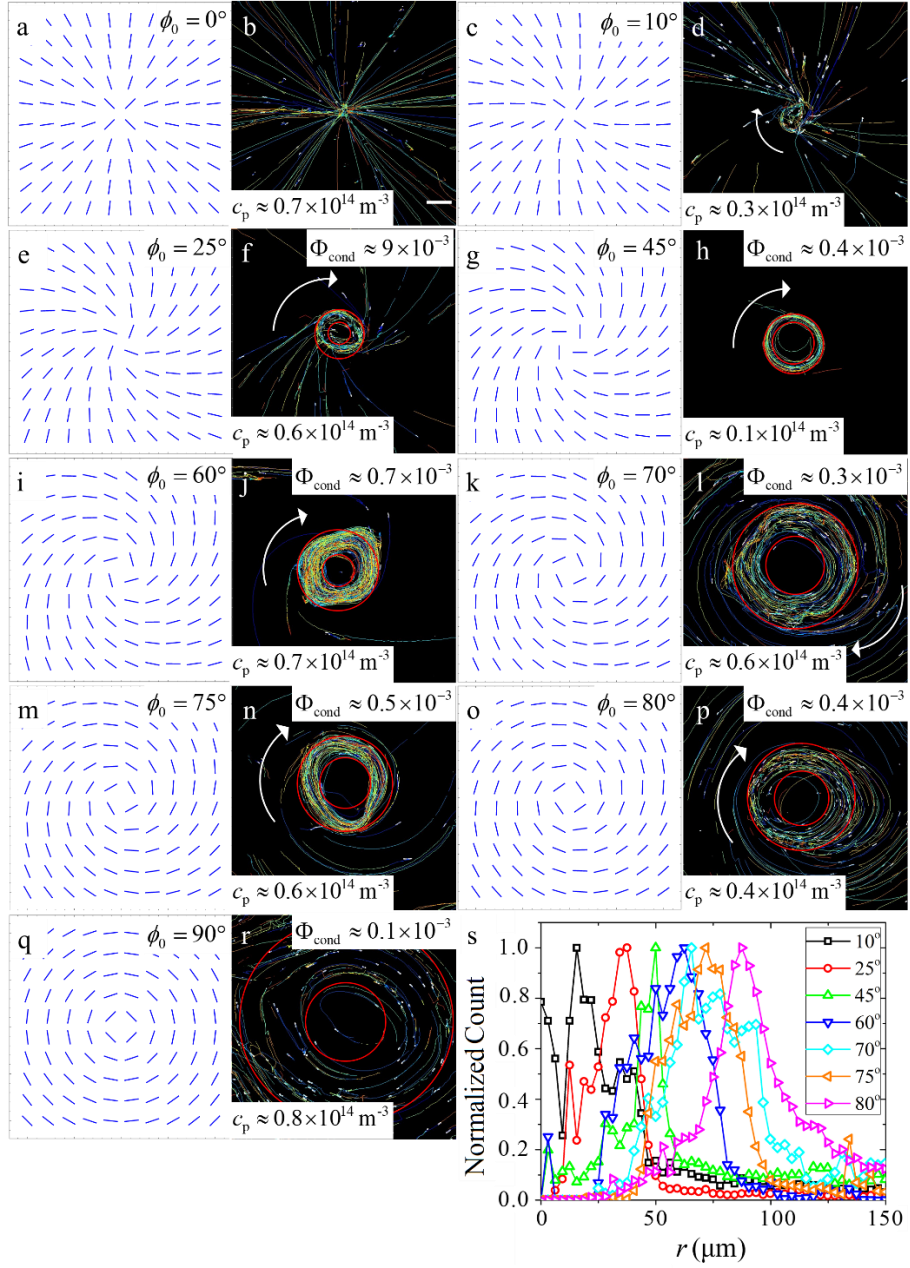

**Figure 2.** Predesigned director fields (left panels) and overlapped trajectories of bacteria in optical microscopy textures (right panels) for (a,b) $\phi_0 = 0$, (c,d) $\phi_0 = 10°$, (e,f) $\phi_0 = 25°$, (g,h) $\phi_0 = 45°$, (i,j) $\phi_0 = 60°$, (k,l) $\phi_0 = 70°$, (m,n) $\phi_0 = 75°$, (o,p) $\phi_0 = 80°$ and (q,r) $\phi_0 = 90°$. Vortices with $0 < \phi_0 < 90°$ produce clockwise circulation of bacterial swirls. All samples are with the same initial concentration $c_v \approx 0.8 \times 10^{13}\,\text{m}^{-3}$, $\Phi_v \approx 2.4 \times 10^{-5}$. The local concentration is in the range

$c_p^{cr} < c < c_p^{max}$. The insets show $c_p$ obtained by averaging instantaneous $c_p$ over all frames within the time interval [0, 60 s]. The trajectories are accumulated over the same time interval. Scale bar 50 μm. (s) Normalized bacterial count per area of torus as a function of the radial distance from the center of each vortex. Radial step length $\Delta r = 3$ μm. The data are averaged throughout [0, 60 s].

A velocity map for each vortex was obtained by tracking each bacterium for 60 s using particle image velocimetry (PIV) MATLAB package [47], Fig. S4a-h [42]. The speed $v = \sqrt{v_r^2 + v_\phi^2}$ is color-coded in Fig. S4. The maps clearly show a clockwise circulation in each vortex. The azimuthal velocity $v_\phi$ changes non-monotonously with the radial distance and $\phi_0$, reaching a maximum $v_\phi \approx 7.2$ μm/s at $\phi_0 = 45°$, and dropping to $v_\phi \approx (6.0 - 6.6)$ μm/s at $\phi_0 = 70°, 80°$ and even to $v_\phi \approx 5$ μm/s at $\phi_0 = 10°, 25°$, Fig.3a. As discussed in the next subsection, a plausible reason is that the azimuthal active force reaches a maximum at $\phi_0 = 45°$. In addition, at $\phi_0 > 45°$, the swirls are wider and less concentrated than the swirls at $\phi_0 = 45°$, while the swirls at $\phi_0 < 45°$ cross $\hat{\mathbf{n}}_0$ at a very large angle and thus need to overcome a stronger viscoelastic resistance [18].

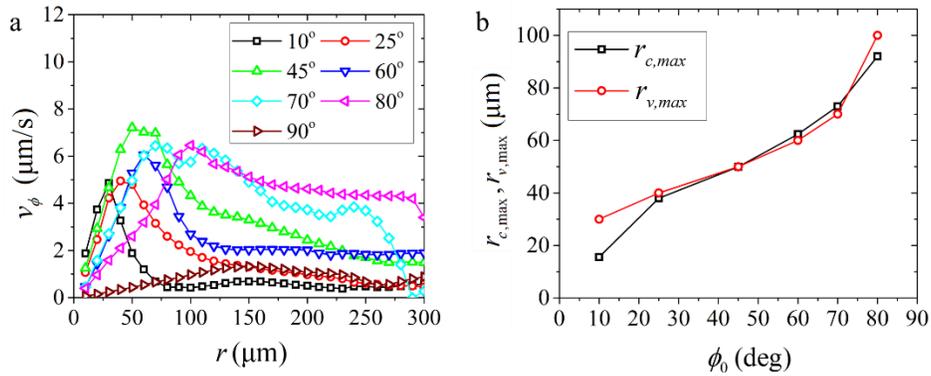

**Figure 3.** (a) Azimuthal component of velocity $v_\phi$ for each pattern as a function of radial direction $r$. (b) Radii $r_{c,max}$ and $r_{v,max}$ as a function of $\phi_0$. All samples are with the same initial concentration

$c_v \approx 0.8 \times 10^{13}$ m$^{-3}$, $\Phi_v \approx 2.4 \times 10^{-5}$. The local concentration of bacteria for all $\phi_0$'s is in the range $0.1 \times 10^{14}$ m$^{-3} \leq c_p \leq 0.8 \times 10^{14}$ m$^{-3}$ ($3 \times 10^{-5} \leq \Phi_p \leq 2.4 \times 10^{-4}$).

This behavior differs from a theoretical prediction [48] that a Taylor swimmer, i.e., a deforming sheet which generates low-amplitude plane waves, placed in a uniform nematic $\hat{\mathbf{n}}_0 = \text{const}$, would swim faster if it were perpendicular to $\hat{\mathbf{n}}_0$ than when it were parallel to $\hat{\mathbf{n}}_0$. In particular, the swimming speed is predicted to increase when the Taylor swimmer makes an angle between 45° and 135° with $\hat{\mathbf{n}}_0$ [48]; this regime corresponds to the splay-rich vortices $0 < \phi_0 < 45°$, but in our case the speed in splay-rich vortices is actually smaller than that at $\phi_0 = 45°$. There are many reasons for the discrepancy: (i) surface anchoring at the bacterial body is strong enough to keep the local director parallel to the bacterial body [18] and thus different from $\hat{\mathbf{n}}_0$, while in the considered model of Taylor swimmer this anchoring is weaker; (ii) the director in our case is pre-patterned, while in Ref. [48] it is uniform; (iii) the modes of propulsion are different, etc. The observed speed dependence on the anisotropy direction of environment demonstrates a rich potential of an orientationally ordered media to control active flows.

The diameter of circular swarms of bacteria increases with the spiral angle $\phi_0$, as clear from Fig.3b, in which each velocity dependency $v_\phi(r)$ peaks at a certain radius $r = r_{\max}$ that increases with $\phi_0$, for a similar duration of observation (60 s).

So far, we described short-term dynamics, mostly within the time interval [0, 60 s]. The circular swirls also show long-term dynamics that is dramatically dependent on $\phi_0$, Fig. 4. Namely, bacterial swirls tend to condense towards the center in the vortices with a prevailing splay, $\phi_0 < 45°$, Fig. 4a-c, j, while they expand and move away to the periphery in the vortices of a predominant bend, $\phi_0 > 45°$, Fig. 4g-h, k. The condensing swirls at $\phi_0 < 45°$ show two types of behavior. In the pure radial geometry, the bacteria could form an immobilized dense cluster with the volume fraction approaching 1, after 5-6 minutes from the start of the experiment. In vortices with $\phi_0 = 10, 25°$, the bacteria continue swimming in the azimuthal direction in the shrunk swirls and do not form dense immobilized clusters, as some bacteria escape the swirls and swim away. The swirls are most stable in the vortices of splay-bend parity, $\phi_0 = 45°$, Fig. 4d-f, where their inner and outer radii remain unchanged for more than 10 min.

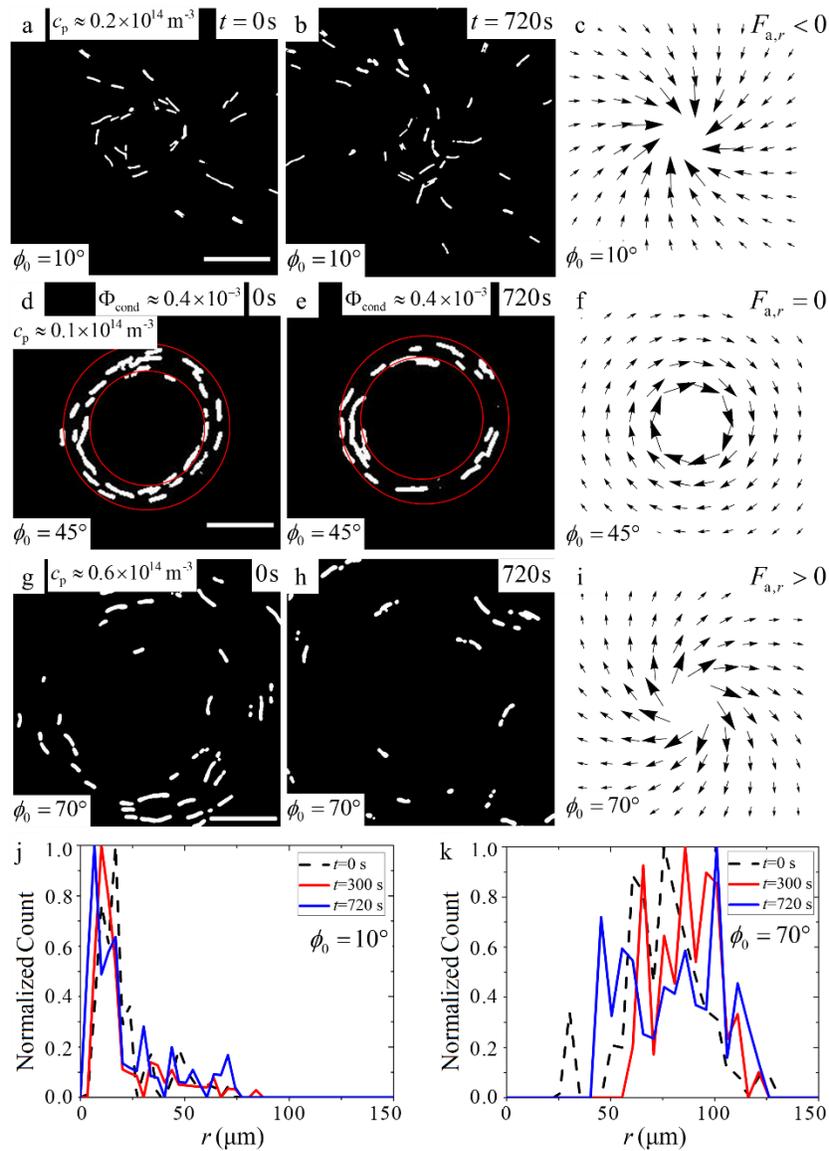

**Figure 4.** Long-term evolution of circulating swirls at different $\phi_0$'s ($c_v \approx 0.8 \times 10^{13}$ m$^{-3}$, $\Phi_v \approx 2.4 \times 10^{-5}$). (a,b) Optical microscopy of contracting swirls in a $\phi_0 = 10°$ vortex, for which (c) the radial component of active force is directed towards the center, $F_{a,r} < 0$. (d,e) Long-term stability of a bacterial swirl at $\phi_0 = 45°$, traced between $t = 0$s and $t = 720$s. (f) The active force at $\phi_0 = 45°$ is strictly azimuthal, directed clockwise, $F_{a,r} = 0$; the azimuthal component reaches its maximum, $F_{a,\phi} = \alpha / r$; (g,h) Expanding swirl at $\phi_0 = 70°$ with (i) centrifugal radial component of the active force, $F_{a,r} > 0$. (j) Normalized distribution of bacteria along the radial direction $r$ for $\phi_0 = 10°$ shown for three instances, $t = 0$s, $300$s, and $720$s. (k) The same for $\phi_0 = 70°$. Scale bar 50 μm.

### 3. *Active force induced by director-mediated hydrodynamic interactions.*

The salient features of the bacterial dynamics in the predesigned vortex director field, namely, onset of collective swirl dynamics, clockwise swimming, shrinking and expansion of swirls, dependence of the azimuthal velocity on the spiral angle can be qualitatively explained by analyzing the active force that emerges when the bacterial force dipoles, being embedded into the passive director field, start to overlap. Figure 1a,d explains the mechanism of the individual-to-collective motion transition for the $\phi_0 = 25°$ vortex. In a general case, the active force as a function of the director gradients can be written in the approximation of a spatially-independent concentration [23,49], $c = \text{const}$, as

$$\mathbf{F}_a = \alpha(\hat{\mathbf{n}}_0 \nabla \cdot \hat{\mathbf{n}}_0 - \hat{\mathbf{n}}_0 \times \nabla \times \hat{\mathbf{n}}_0), \tag{5}$$

where $\alpha = -\lambda c$ is the activity coefficient defined as the product of bacterial concentration and force dipole $\lambda > 0$ exerted by each bacterium on the surrounding nematic; $\alpha$ is negative since *B. subtilis* microswimmers are pushers [50,51]. The limitation $c = \text{const}$ is lifted in the next section, which operates with the full description of the active force as a divergence of the active stress. Note that in strongly condensed systems, such as a condensed dispersion of filaments, $\alpha$ is often taken as proportional to $c^2$, to stress that the activity arises from pairwise interactions [52]. In our case, the active force in a relatively dilute subsystem of bacteria is mediated by the passive nematic environment, so that $\alpha \sim c$ to reflect its superposition character.

For the director field in Eq. (1), which spirals counterclockwise when $0 < \phi_0 < 90°$, the azimuthal component of the active force in Eq. (5), $F_{a,\phi} = \frac{\alpha}{r}\sin 2\phi_0$, is always negative (since $\alpha < 0$), i.e., directed clockwise, Fig.1d, Fig.4c,f,i. This negative azimuthal component coerces the bacteria to swim clockwise in swirls for any $0 < \phi_0 < 90°$. In contrast, the radial component $F_{a,r} = \frac{\alpha}{r}\cos 2\phi_0$ can be either negative, $\phi_0 < 45°$, Fig. 4c (compare to Fig.1a), or positive, $\phi_0 > 45°$, Fig.4i, thus facilitating shrinking or expansion of the swirls, respectively. The radial component is zero when $\phi_0 = 45°$, Fig.4f. These qualitative geometrical arguments explain the experimental scenarios: the swirls circulate clockwise and either shrink to the center when $\phi_0 < 45°$, or migrate to the periphery when $\phi_0 > 45°$, Fig. 4 and Supplemental Material Fig. S3 [42]. The vortex with $\phi_0 = 45°$ produces the most stable swirls, since the active force is strictly azimuthal, with zero radial component, Fig.4d-f. The azimuthal component of the actice force, $F_{a,\phi} = \frac{\alpha}{r}\sin 2\phi_0$, attains its maximum absolute value at $\phi_0 = 45°$, which helps the azimuthal velocity $v_\phi$ to reach its maximum in vortices with $\phi_0 = 45°$, Fig.3a.

As demonstrated by Zhou et al [18] for a different geometry of homeotropic samples, swimming bacteria can realign the director field parallel to their rod-like bodies, even if the surface anchoring at the cell's substrates sets the director orientation different from the orientation of the bacterial body. A similar effect occurs in circular swirls. Realignment of the director from $\hat{\mathbf{n}}_0$ is easier to observe in thin rather than in thick cells since one can use PolScope to map the actual director. We used cells with $d \leq 6.5\,\mu\text{m}$ and $\phi_0 = 45°$. Figs. 5 a,b clearly demonstrate that the director realigns from the pre-imposed direction towards the circular trajectories of the swimming bacteria. The angle $\beta$ between the actual director field in the swirl and the predesigned director $\hat{\mathbf{n}}_0\,(\phi_0 = 45°)$ is significantly different from zero, with an average value $\overline{\beta} \approx 24°$, Fig. 5b.

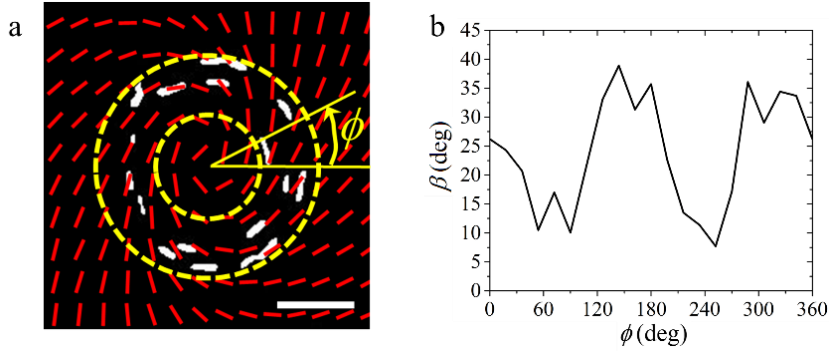

**Figure 5.** Swirling-induced director deviations from surface-imposed $\hat{\mathbf{n}}_0$ in a spiral vortex $\phi_0 = 45°$; data accumulated over the interval [0, 60 s]. (a) Optical microscopy texture of bacteria (white) in a steady swirl (confined between two yellow dashed circles) overlapped with the actual director (red), determined by PolScope. Scale bar $50\,\mu\text{m}$. (b) Azimuthal variations of the angle $\beta$ between the actual director field in the bacterial swirl and $\hat{\mathbf{n}}_0$; $\beta$ is calculated by dividing the swirl into segments of $\Delta\phi = 18°$ and finding the angle in each segment.

To conclude this section, it is important to stress that the collective unipolar swirling occurs only for the mixed splay-bend deformations, $0 < \phi_0 < 90°$. Vortices of pure splay and pure bend produce no net flows. In the splay case ($\phi_0 = 0$), the bacteria swim in and out of the vortex core along the radial directions, Fig.2a and Supplemental Material Fig.S2c [42], but when the concentration is raised to $c_v \approx 0.16\times 10^{14}\,\text{m}^{-3}$, $\Phi_v \approx 0.5\times 10^{-4}$, they accumulate in the center [23], Supplemental Material Fig. S2d [42], forming an immobilized dense cluster within a few minutes.

In the bend case ($\phi_0 = 90°$), the bacteria swim in a bipolar fashion along circular trajectories that gradually displace to the periphery, see Supplemental Material Fig.S2b [42] and Fig.2r. The absence of net flows, condensation at the cores, and expansion to the periphery are explained by an argument similar to the scheme in Fig.1a and by Eq. (5). Namely, the active force $\mathbf{F}_a$ in these cases is purely radial, either centripetal, $\phi_0 = 0$, or centrifugal, $\phi_0 = 90°$, with no azimuthal component.

### 4. *Spiral vortices: regime of undulations and population oscillations.*

The circulating swirls deviate from a circle to an undulating shape, Fig.6, when the number of bacteria increases above some threshold $c_p^{max}$ that depends on $\phi_0$. In order to observe the undulations, we use a higher initial concentration $c_v \approx 0.1 \times 10^{14}$ m$^{-3}$, $\Phi_v \approx 3 \times 10^{-5}$. The undulations arise from bending instability that is typical of the dynamics of "pushers" [17,36]. Figure 6 illustrates the development in a $\phi_0 = 45°$ vortex. Once the local concetration raises above $c_p \approx 0.1 \times 10^{14}$ m$^{-3}$, the bacteria form a steady circular swirl with a local condensation volume fraction $\Phi_{cond} \approx 0.6 \times 10^{-3}$, Fig.6a. As time goes by, the swarm attracts more bacteria, Fig.6b, their concentration increases to $c_p^{max} \approx 1.0 \times 10^{14}$ m$^{-3}$, $\Phi_{cond}^{max} \approx 0.8 \times 10^{-3}$, at which point the circular shape begins to undulate, Fig. 6c. Once the swirl starts to undulate, the bacteria escape from it, through openings in the most developed protrusions, Fig.6d,e. After releasing some bacteria, the swarm becomes less populated, Fig.6f, with the concentration dropping down to $c_p \approx 0.3 \times 10^{14}$ m$^{-3}$, $\Phi_{cond} \approx 0.6 \times 10^{-3}$, which is below the undulation threshold $c_p^{max}$, $\Phi_{cond}^{max} \approx 0.8 \times 10^{-3}$, Fig. 6g. As a result, the swarm restores its circular closed shape, resuming steady swirling, Fig. 6h. The circulation remains steady until the concentration increases to $c_p \approx c_p^{max} \approx 1.0 \times 10^{14}$ m$^{-3}$, $\Phi_{cond}^{max} \approx 0.8 \times 10^{-3}$ again, after which the cycle of undulations and population oscillation repeats itself. The entire cycle can be repeated multiple times.

Similar undulations and population oscillations can also be observed for other vortices with $\phi_0 > 45°$, as illustrated for $\phi_0 = 75°$ in Supplemental Material Fig.S5 [42]. However, after the initial undulation and release of some bacteria, the swarms form circular swirls with a radius larger than the initial one. The reason is that the swirls are expanded by the outward radial component of the active force that is present at any $\phi_0 > 45°$. For $\phi_0 < 45°$, condensation towards the center occurs before the undulations could develop.

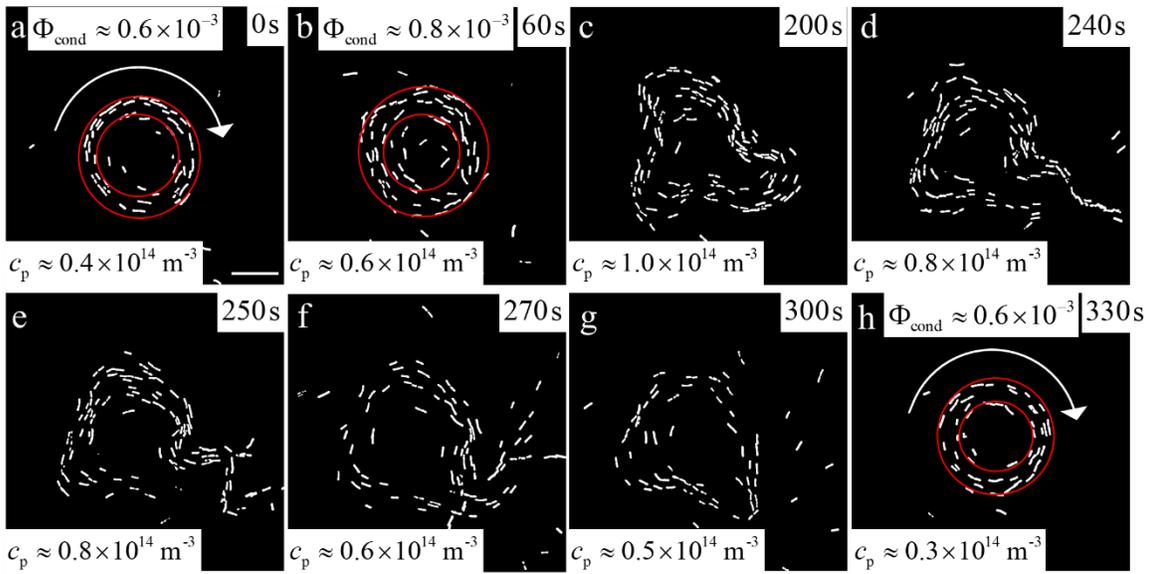

**Figure 6.** Optical microscopy textures of time sequence of cyclic population oscillation through undulations and partial release of bacteria in $\phi_0 = 45°$ swirls; the sample is prepared with a high initial concentration $c_v \approx 0.1 \times 10^{14}\,\text{m}^{-3}$, $\Phi_v \approx 3 \times 10^{-5}$, to achieve the undulation threshold. (a) Steady circulation. (b) Increase of number of bacteria in the swirl. (c) Undulations. (d) Rupture of the swirl at the most developed protrusions. (e,f) Escape of some bacteria from the swarm. (g) Healing of the circular swirl by closure. (h) Circular swirling is restored. Scale bar $50\,\mu\text{m}$.

It is of interest to compare the undulation threshold $\Phi_{\text{cond}}^{\max} \approx 0.8 \times 10^{-3}$ with a similar threshold volume fractions that lead to undulations in other geometries of the director field. In uniformly aligned samples, the unulations occur at noticeably lower concentrations

$\Phi_v = \Phi_p = \Phi_{cond} = 0.3 \times 10^{-3}$ ($c_v = c_p \approx 1.0 \times 10^{14}$ m$^{-3}$), see Supplemental Material Fig. S1b [42]. On the other hand, this $\Phi_{cond}^{max} \approx 0.8 \times 10^{-3}$ is much smaller than the threshold concentration $\Phi_p \approx \Phi_{cond} = 3.6 \times 10^{-3}$ ($c_p = 11.8 \times 10^{14}$ m$^{-3}$) in one-dimensional "C"-patterns with alternating splay and bend stripes [25]. In other words, the vortices of the background passive director help to stabilize the collective unipolar motion as compared to the uniform director environment, but not as strongly as the C-patterns of alternating bend and splay. These differences illustrate that the undulation threshold is different in different geometries, a fact that can be used to stabilize or destabilize the dynamics of bacteria by a suitable design of the director field.

5. *<u>Numerical simulations of the individual-to-collective motion: agent-based approach</u>*.

The experiment, Fig.1, demonstrates that the transition from an individual bidirectional swimming to a collective unipolar circular swirling occurs at a relatively small number of microorganisms, on the order of 10. The effect of the spiral angle on swirls is also evident for relatively small number of bacteria involved, Figs. 4,5. These data suggests that in order to capture best these striking effects, one should use a model based on discrete active units. These units interact through the underlying passive director pattern $\hat{\mathbf{n}}_0(\mathbf{r})$. The director tends to align the dispersed rod-like bacteria along itself because of the tangential anchoring of the director at the bacteria's bodies [17,18].

We simulate individual and collective dynamics of a finite number $n_B$ of swimming bacteria in the 2D prepatterned passive director field $\hat{\mathbf{n}}_0 = (\cos\phi_0, \sin\phi_0)$. The goal is to explore whether the agent-based model could describe a transition from the nonpolar individual to collective unipolar circular swimming in a system with finite number of bacteria that interact through the underlying director pattern. Each bacterium is specified by its position $\mathbf{r} = (x, y)$ and velocity $\hat{\mathbf{p}}V_0$, where $\hat{\mathbf{p}} = (\cos\theta, \sin\theta)$ is the unit vector parallel to the long bacterial axis, $\theta$ is the angle between the $x$-axis and the vector $\hat{\mathbf{p}}$. We use a nematic type of interactions between the bacteria and the passive director, so that the evolution of the bacterial orientation $\theta$ with respect to the director angle $\phi_0$ is described by the equation $\partial_t \theta = \gamma_B \sin\left[2(\phi_0 - \theta)\right]$, where $\gamma_B$ is the relaxation rate that depends on the surface anchoring strength at the bacterial body ($\gamma_B \to \infty$) when the bacteria follow the local director at any moment of time); factor 2 reflects the nematic nature of interactions, in which the outcome does not depend on a realignment of $\theta$ by $\pi$. In the vector form, this relaxation dynamics equation for the bacterial orientation writes $\partial_t \hat{\mathbf{p}} = \gamma_B \hat{\mathbf{n}} \times \hat{\mathbf{p}}(\hat{\mathbf{n}}\hat{\mathbf{p}})$. Each individual bacterium exerts an active stress $\boldsymbol{\sigma}_{act} = -\lambda \mathbf{Q}_B f(\mathbf{r})$ onto the nematic surrounding, where $\lambda$ is the dipolar strength of a bacterium (unit of length is normalized by the length of a bacterium; the unit of velocity is normalized by the bacterial swimming speed; and the unit of force is normalized by the bacterial propulsion force); the "-" sign corresponds to the pusher nature of the swimmers, $\mathbf{Q}_B = \hat{\mathbf{p}}\hat{\mathbf{p}} - \mathbf{I}/2$ is the tensorial order parameter of the bacteria, $\mathbf{I}$ is the identity tensor, and $f(\mathbf{r})$ is anisotropic Gaussian distribution which models the anisotropic shape of a bacterium:

$$f(\mathbf{r}) = \exp\left[-(x\cos\theta + y\sin\theta)^2/g_\parallel^2 - (-x\sin\theta + y\cos\theta)^2/g_\perp^2\right]/(2\pi g_\parallel g_\perp),$$

with the parameters $g_\parallel = 2.2l$ and $g_\perp = 1.0l$ that define the width of the distribution along $\hat{\mathbf{p}}$ and perpendicular to $\hat{\mathbf{p}}$, respectively. The active stress generates a realigning force $\mathbf{F}_{act} = \nabla \cdot \sum \boldsymbol{\sigma}_{act}$, where the sum is taken over all bacteria. A relaxation equation in the Landau-Lifshitz-Gilbert form for the director in the bulk writes $\partial_t \hat{\mathbf{n}} = -\hat{\mathbf{n}} \times \hat{\mathbf{n}} \times (\gamma_s \hat{\mathbf{n}}_0 - 2a\mathbf{F}_{act} + K\Delta\hat{\mathbf{n}})$, where $\gamma_s$ is the director relaxation rate dependent on the director anchoring strength at the bounding plates and $a$ is a coefficient dependent on the surface anchoring strength at the bacterial body [18], $K$ is the Frank elastic constant of the nematic normalized by rotational friction (for more details, see Supplemental Material [42]). This form automatically ensures the director normalization condition $\hat{\mathbf{n}}\hat{\mathbf{n}}=1$. The Landau-Lifshitz-Gilbert dynamic equation is fully equivalent to the Leslie-Ericksen description of the director realignments but it is easier to use in numerical simulations since it does not encounter the angle discontinuity intrinsic to the nematic angle description; it can be applied as long as the director field does not contain half-integer disclinations, which is the case under consideration. Note that our model does not include bacterial interactions among themselves nor the effects of random noise, because of the following reasons. The multi-body bacterial interactions are important in dense systems in which the microorganisms are in constant contact with each other, see, for example, Li et al [53]. Our system is much more dilute, with the volume fractions even in the condensed swirls being less than 0.01. Random reorientations of the bacteria are suppressed by the surrounding passive director field and finite anchoring of the director at the bacterial bodies [17,18,31].

Numerical simulations reproduce the most salient feature of the bacterial dynamics in vortices, namely, a transition from the individual swimming parallel to the pre-imposed director $\hat{\mathbf{n}}_0$ at small concentrations to a collective circular swirling above some concentration threshold. In the numerical simulations for $\phi_0 = 45°$ vortex, we use $\gamma_B = 0.2\,\text{s}^{-1}$, $\gamma_S = 0.1\,\text{s}^{-1}$ and $\lambda a \approx 0.5\,\text{s}^{-1}$. We vary the number of individual bacteria $n_B$ from 4 to 256 and find that the collective swirling starts at $n_B \sim 10$, Figure 7 and Supplemental Movie S5 [42], which is practically the same as the experimental one.

The agent-based model captures remarkably well the underlying physics of the individual-to-collective motion transition in the $\phi_0 = 45°$ vortices. At very low concentrations, $n_B = 4, 9$ in Fig.7 and Supplemental Movie S5 [42], the surface anchoring and elasticity of the passive nematic environment are sufficient to align and guide the swimming bacteria strictly along $\hat{\mathbf{n}}_0$, along bipolar trajectories, as in Fig.1b. Once $n_B$ increases above 10 in Fig.7 and Supplemental Movie S5 [42], the hydrodynamic force dipoles of neighbouring bacteria that are aligned by the spiraling $\hat{\mathbf{n}}_0$ at some angle to each other, start to overlap and produce an active azimuthal force that sets a collective polar circular swimming. At this stage, the trajectories are no longer along the director and cut the substrate-imposed $\hat{\mathbf{n}}_0(\mathbf{r})$ at $45°$.

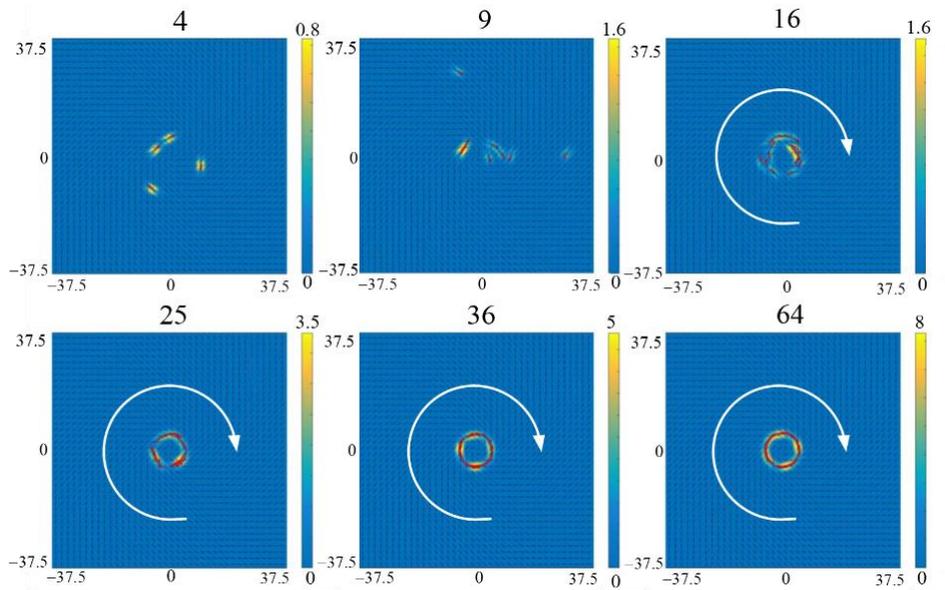

**Figure 7.** The individual-to-collective swimming transition in a the $\phi_0 = 45°$ vortex, as simulated in the agent-based model for a different number $n_B$, ranging from 4 to 64, as indicated. The snapshots for $n_B = 4, 9$ show individual dynamics with bacteria swimming along the spiraling nonpolar director (black ticks). For higher concentrations, $n_B = 16, 25, 36, 64$ the smimming is collective: active units form stable circular swirls circulating clockwise. Bacteria are shown as red arrows. Color coding represents anisotropic Gaussian force exerted by the bacteria.

6. *<u>Numerical simulations of the effect of the spiral angle: agent-based approach</u>.*

The agent-based model also describes the effect of the spiral angle in the range $40° \leq \phi_0 < 90°$ on the bacterial swirls, as illustrated for $n_B = 256$ in Fig. 8 and Supplemental Movie S6 [42]. In the circular vortices, $\phi_0 = 90°$, the bacterial swirl expands with time, Fig. 8g, which is consisted with the experiment and the qualitative active force description, since the radial outward component of the active force is maximum at $\phi_0 = 90°$. The radii of toroidal swirls, determined at the maximum simulation time $T$, increase with $\phi_0$, again in agreement with the experiment, Fig. 3b.

Despite the successful description of the individual-to-collective transition and dependence of swirls dynamics on $\phi_0$, we also find certain disagreements between the model and the experiments. Swirls around vortices $40° \leq \phi_0 < 90°$ show only a minor change with time at the late stages of simulations, Fig. 8g. Once formed, the swirls remain stable and show no undulations as long as the simulations run. Furthermore, in vortices with $\phi_0 < 40°$, the simulated microswimmers follow the director $\hat{\mathbf{n}}_0$ and do not form closed circular swirls, even if their number is as high as 900. The most probable reasons for the discrepancy are that the simulations (i) neglect the finite size and steric interactions of the microswimmers and (ii) do not capture fully the length scale of hydrodynamic interactions of bacteria and the complex pattern of surface anchoring at their bodies (which is especially important at $\phi_0 < 45°$ since in this case the rod-like bacterium in a swirl forms a large angle $\frac{\pi}{2} - \phi_0$ with the pre-patterned director).

Further insights are obtained in the continuum (mean-field) analytical approximation of the agent-based model, see Supplemental Material [42]. In this approximation, one derives an equation for the concentration of bacteria and the director $\hat{\mathbf{n}}$; $\hat{\mathbf{n}}$ is generally different from the prescribed $\hat{\mathbf{n}}_0$ because of the torques and forces imposed by bacteria. Neglecting the elastic response of the nematic, one obtains an analytical expression for the director $\hat{\mathbf{n}}$ vs concentration, Eq. (S11). The resulting relation is used to solve the problem analytically. The steady-state distributions of concentration vs the radius $r$ for different values of the prepatterned spiral angle are shown in Supplemental Material Fig. S6 [42]. In faithful agreement with the agent-based model, one obtains a monotonic increase of the swirl radius with the concentration and the angle $\phi_0$, Fig. 8h.

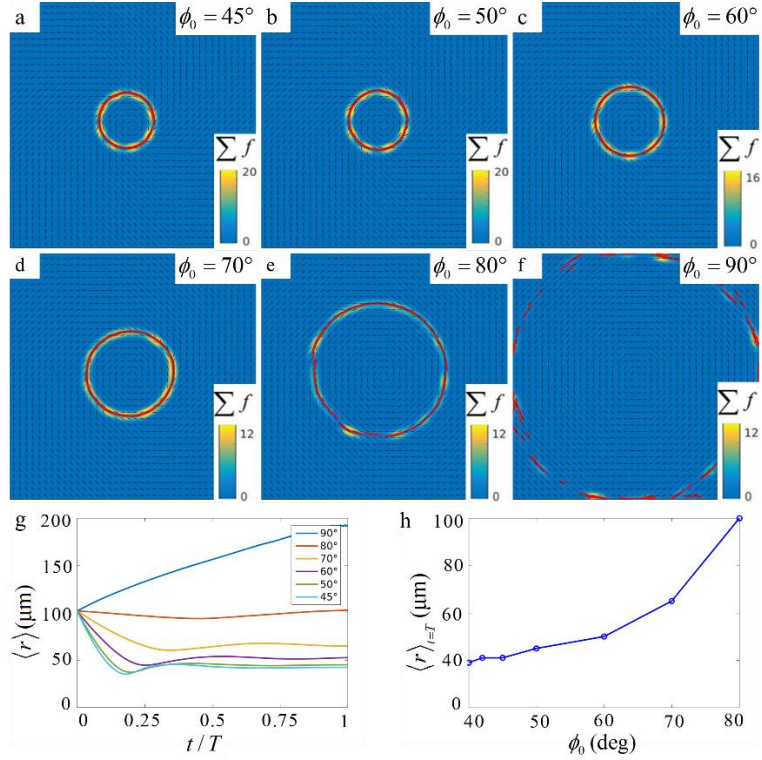

**Figure 8.** Simulation of the agent-based model. Stable circulations settle around vortices with (a) $\phi_0 = 45°$, (b) $\phi_0 = 50°$, (c) $\phi_0 = 60°$, (d) $\phi_0 = 70°$, (e) $\phi_0 = 80°$, (f) $\phi_0 = 90°$ for $n_B = 256$ at maximum simulation time $T$. Bacteria are shown as red arrows. Color coding in (a-f) represents anisotropic Gaussian force exerted by the bacteria. (g) The average distance from the center of vortices as the function of simulation time. (h) Radius of the circulating swirl as the function of $\phi_0$

## 7. *Numerical simulations of the bacterial swirls: continuum approach.*

Despite its simplicity and clear underlying physical mechanisms, the agent-based model fails to capture the behavior of relatively dense swirls that show undulatory instabilities, Fig.6. To explore this part of the rich dynamic spectrum of the bacterial dynamics in living liquid crystals, we employ the advection-diffusion model [25], see Eqs. (2)-(4) in the Materials and Methods section below. The advection-diffusion model directly incorporates the fluid velocity and makes an experimentally tested prediction that this velocity in stable swirls is directed clockwise parallel to the bacterial velocity, thus helping the swirl's circulation. Furthermore, the advection-diffusion model successfully captures the onset of undulations of condensed swirls, Fig.6, as demonstrated below.

We numerically integrated Eqs. (2)-(4), on graphical processor units (GPUs), see Ref. [36] for numerical implementation details. Starting with random initial conditions, the system self-organizes and exhibits behavior similar to that observed in the experiment: when the concertation exceeds some threshold, bacteria form a circular swirl. In the swirls, all bacteria swim unidirectionally and create unidirectional fluid flows, Fig. 9a,b. This behavior is consistent within the range $\phi_0 \in [65°, 85°]$, Fig. 9c. Furthermore, we observe a monotonous decrease of the swirl radius while decreasing $\phi_0$ towards $45°$.

At higher bacterial concentrations, the circular trajectories start to undulate, Fig. 9d and Supplemental Movie S7 [42]. In agreement with the experiment, bacterial distribution in undulating swirls is non-uniform along the swirl arclength, Fig. 9d. As in the agent-based simulations, we do not observe persistent circular collective motion for small $\phi_0$, as in these vortices, the model shows that the bacterial concentration reaches its maximum in the vortex center. This is likely due to the limitations discussed in the previous section on agent-based approach and because of the oversimplified expression for the active stress $\boldsymbol{\sigma}_{\text{act}} \sim \mathbf{Q}_B$. In the last expression, the radial part of the active force changes sign at $\phi_0 = 45°$, which leads to accumulation of bacteria at the origin for $\phi_0 < 45°$. Higher-order corrections to $\boldsymbol{\sigma}_{act}$ should improve the agreement.

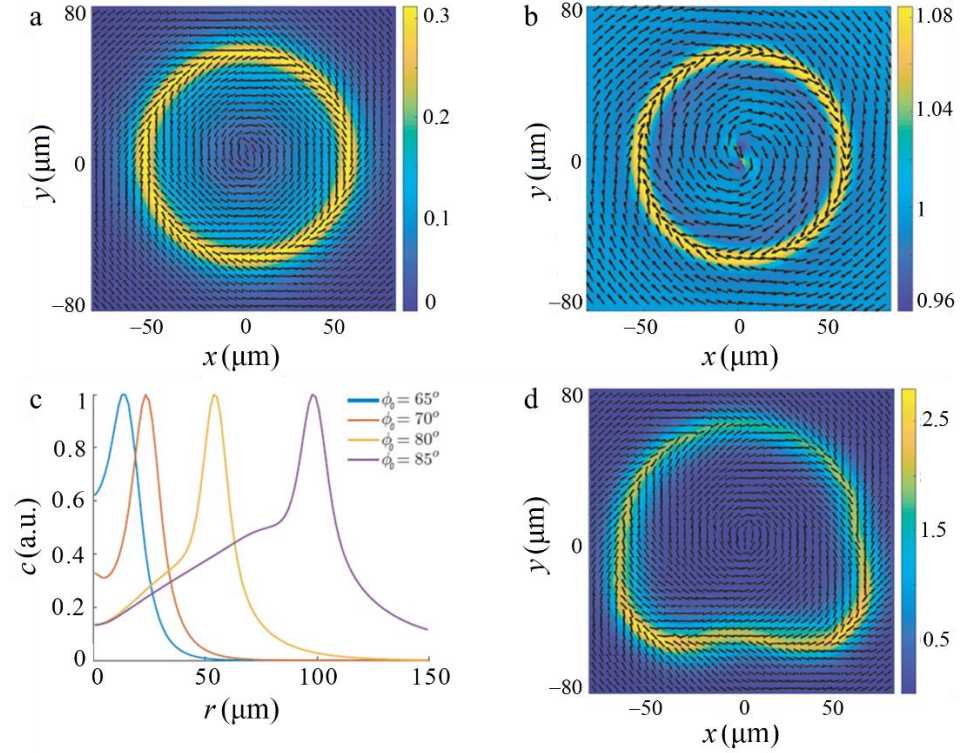

**Figure 9.** Simulation results of full advection-diffusion model Eqs. (2)-(4). (a) Bacterial concentration (colormap) and nematic field (lines). (b) Full velocity field (colormap and arrows) is a superposition of bacterial self-propulsion speed (magnitude normalized by 1) and hydrodynamic velocity. Panels (a,b) show the regime of stable circulation at $\phi_0 = 80°$. (c) The dependence of bacterial concentration (maximum value normalized by 1) on distance from the vortex center for various $\phi_0$'s. (d) Bacterial concentration (colormap) and nematic field (lines) for the regime with undulations.

## IV. DISCUSSION AND CONCLUSION

In this work, we demonstrate the transition between an individual and collective modes of motion of bacteria in a spiral director field of a passive nematic background. In the individual regime, bacteria simply follow the pre-inscribed director, remaining parallel to it. This regime is observed when the concentration of bacteria is small, i.e., the separation distance between bacteria is considerably larger than the range of hydrodynamic interactions, which is on the order of a few tens of micrometers (the typical length of fluid perturbations in the form of hydrodynamic force dipoles produced by an individual motile bacterium in the LCLC [31]). Once the concentration exceeds some threshold $c_p^{cr} \approx 0.7 \times 10^{13}$ m$^{-3}$, which is still a very dilute regime, $\Phi_p^{cr} \approx 2.1 \times 10^{-5}$, the bacteria placed in the spiral vortices ($0 < \phi_0 < 90°$) with the director spiraling counterclockwise, begin to swim collectively, forming swirls that circulate around the vortex center in a clockwise manner. The transition from individual apolar swimming with no net flow to collective unipolar circulation with a net flow is triggered by interactions of bacterial force dipoles in close proximity to each other, mediated by the underlying passive director. The overlap of their hydrodynamic force dipoles, aligned at an angle to each other because of the underlying director pattern, produces an azimuthally- and radially-resolved active force illustrated schematically in Figs.1a,d, 4c,f,i and detailed in the proposed models and simulations. For any director that spirals counterclockwise, $0 < \phi_0 < 90°$, there is always a nonzero azimuthal component $F_{a,\phi} = \frac{\alpha}{r} \sin 2\phi_0$ of the active force [49] in Eq. (5), that coerces the bacteria to swim clockwise. The unipolar circulation is absent in radial vortices, $\phi_0 = 0$, in which the bacteria accumulate at the core, and in circular vortices, $\phi_0 = 90°$, where the circulation is bipolar. The active force in these two cases is purely radial, $F_{a,r} = \pm \alpha / r$, being centripetal ($\phi_0 = 0$, $F_{a,r} = \alpha / r < 0$) or centrifugal ($\phi_0 = 90°$, $F_{a,r} = -\alpha / r > 0$), respectively, with no azimuthal component.

The described outcome of bacterial hydrodynamic interactions mediated by the underlying passive director is dramatically different from the case of an isotropic environment. In an isotropic fluid, such as water, the interactions of hydrodynamic force dipoles prevent stable parallel swimming of bacteria and produce turbulent-like incoherent flows [34,35]. In contrast, surface anchoring and orientational elasticity of the nematic streamline the interactions of the hydrodynamic force dipoles into coherent dynamic patterns. Although the transition from an individual to collective swimming is associated with realignment of the bacterial bodies from being parallel to $\hat{\mathbf{n}}_0$ to making an angle $\frac{\pi}{2} - \phi_0$ with $\hat{\mathbf{n}}_0$, this realignment is steady, as the bacterial swirls maintain their circular shape in a broad range of concentrations and spiral angles.

Swirls' radii for a given concentration of bacteria and for the same moment of time increase monotonously with the spiral angle $\phi_0$. As a function of $\phi_0$ and time, the swirls demonstrate three profoundly different long-term scenarios. In the splay-dominated vortices, $\phi_0 < 45°$, as time goes by, the swirls shrink towards the center. In the bend-dominated vortices, $\phi_0 > 45°$, the swirls expand to the periphery. The collective circular swirling is most stable in vortices with $\phi_0 = 45°$, where it persists unchanged for a steady level of bacterial activity. These vortices also show the highest speed of collective circulation, which we associate with the fact that the azimuthal component of the active force $F_{a,\phi} = \frac{\alpha}{r} \sin 2\phi_0$ reaches its extremum at $\phi_0 = 45°$.

Formation of swirls in any $0 < \phi_0 < 90°$ vortex is reminiscent of a stable limit cycle. If one considers the entire set of vortices, with the spiral angle as a parameter, then this set singles the $\phi_0 = 45°$ vortex as an example of an unstable limit cycle: an increase of $\phi_0$ expels the bacterial trajectories to the periphery, while a decrease of $\phi_0$ moves the trajectories closer to the center, in both cases departing the stable circulation with a constant radius.

The differences in expansion/contraction is caused by the sign change of the radial component of the active force, $F_{a,r} = \frac{\alpha}{r}\cos 2\phi_0$. When $\phi_0 < 45°$, then $F_{a,r} < 0$ which causes contraction of swirls and condensation of bacteria over time. When $\phi_0 > 45°$, the radial force is centrifugal, $F_{a,r} > 0$, causing an expansion of swirls. When $\phi_0 = 45°$, there is no net radial force, so the radius of the bacterial swirl remains constant.

The bacterial concentration impacts their collective dynamics very strongly. Steady circulation occurs only within a limited concentration range $c_p^{cr} < c < c_p^{max}$ ($c_p^{cr} \approx 0.7 \times 10^{13}$ m$^{-3}$ and $c_p^{max} \approx 10 \times 10^{13}$ m$^{-3}$), when the volume fraction of bacteria is above $\Phi_p^{cr} \approx 2.1 \times 10^{-5}$ but below its maximum value calculated within the volume occupied by the swirl, $\Phi_{cond}^{max} \approx 0.8 \times 10^{-3}$. Below this range, the bacteria swim individually, following apolar trajectories parallel to the passive spiral director and creating no net flow. Above this range, the circular bacterial swirls exhibit undulations. The undulating swirls can heal themselves back to the circular shape by opening up and releasing some bacteria, thus reducing the concentration below $c_p^{max}$. In $\phi_0 = 45°$ vortices, the cycle of population oscillations through steady circulation -> undulation -> bacterial release->healing->circulation can repeat itself multiple times.

The experimental findings for the short-time dynamics of the bacterial swirls are well described by the theoretical models. Some disagreement for long-time behavior and swimming in vortices with a small $\phi_0$ can be attributed to the limitations of the theory, such as the omission of steric interactions between the bacteria in condensed swirls and relatively small integration domains. Large angular discrepancies between the alignment of the bacterial bodies and the prepatterned $\hat{\mathbf{n}}_0$ add to the complexity of the problem. The long term behavior is also influenced by the experimental factors that make modeling harder, such as decrease of activity with time due to the depletion of nutrients.

Our results stipulate new design concepts on how to harness individual and collective motion of microswimmers, both living and synthetic, to extract a useful mechanical work that might power microscopic systems. Several realizations of active matter powered microdevices were proposed recently in Refs. [10,11,54]. As we have demonstrated, a liquid crystalline environment with patterned surface alignment enables self-concentration of bacteria in predesigned locations, without hindering the space with impenetrable walls, thus facilitating energy transduction. Note that the liquid crystal under consideration is a non-toxic aqueous solution of disodium cromoglycate (used as an inhaled anti-inflammatory agent for the treatment of asthma), in which water occupies about 86% of volume. Such a material can be easily interfaced with other microfluidic elements. Of course, bacteria can be replaced with artificial swimmers, which would expand the spectrum of potential liquid crystalline environments for controlled microdynamics.

**Acknowledgments**
The authors are thankful to M. Carme Calderer for pointing limit cycle analogy, A. Doostmohammadi, R. Green, D. Golovaty, K. Thijssen, J. Toner, V. Vitelli, N. Walkington, J.M. Yeomans, and participants of UC Santa Barbara Kavli Institute for Theoretical Physics (KITP) program "*Active 20: Symmetry, Thermodynamics, and Topology in Active Matter*" for fruitful discussions. This work was supported by NSF grants DMS-1729509 (analysis of bacterial dynamics), CMMI-1436565 (preparation of plasmonic metamasks), REU program CHE-1659571 at Kent State University (support of RJL, a REU student) and by the Office of Sciences, DOE, grant

DE-SC0019105 (development of patterned liquid crystal elastomer coatings). This research was completed while RK and ODL participated in KITP Active 20 program, supported in part by the NSF grant PHY-1748958 and NIH grant R25GM067110. The work of ISA was supported by the NSF grant PHY-1707900.

**Data availability**

The data in support of the reported findings and computer code are available from the corresponding author upon request.

## Supplementary Information

**Swimming bacteria in a uniform director field.** In order to establish the reference state, we explore the dynamics of bacteria in uniformly aligned planar cells of a thickness $d = 20\,\mu\text{m}$, as a function of their overall concentration $c_v$, Fig. S1a-c; $c_v$ is calculated as the number of bacteria per unit volume, which does not depend on the director pattern. At low concentrations, $c_v \approx 0.3 \times 10^{14}\,\text{m}^{-3}$ (volume fraction $\Phi_v \approx 0.9 \times 10^{-4}$), the average speed of bacteria is $\langle v \rangle \approx 8.0\,\mu\text{m/s}$, Fig. S1a. When $c_v$ increases to $\approx 1.0 \times 10^{14}\,\text{m}^{-3}$, $\Phi_v \approx 3 \times 10^{-4}$, bacterial trajectories begin to undulate and acquire nearly periodic bend with a wavelength of $\xi \approx 160\,\mu\text{m}$, Fig. S1b, while their average speed reaches $\langle v \rangle \approx 9.4\,\mu\text{m/s}$. At even higher concentration $c_v \approx 3.1 \times 10^{14}\,\text{m}^{-3}$, $\Phi_v \approx 9.3 \times 10^{-4}$, the wavelength of bending shortens to $\xi \approx 140\,\mu\text{m}$, Fig. S1c, and the average speed increases to $\langle v \rangle \approx 16.6\,\mu\text{m/s}$. Thus, the average speed of bacteria increases with concentration. This trend is in agreement with the study by Sokolov et al [1] of the concentration dependence of the bacterial velocity in isotropic aqueous medium. The second conclusion is that the concentration $\approx 1.0 \times 10^{14}\,\text{m}^{-3}$, $\Phi_v \approx 3 \times 10^{-4}$, in cells with $d = 20\,\mu\text{m}$ is a critical concentration at which the bacterial dynamics becomes unstable with respect to bending instabilities; this result is in good agreement with the data by Zhou et al [2].

**Radial vortex (aster).** In the case of splay, $\phi_0 = 0$, in moderately concentrated dispersions, $c_v \approx 0.8 \times 10^{13}\,\text{m}^{-3}$, $\Phi_v \approx 2.4 \times 10^{-5}$, the bacteria swim in and out of the vortex core along the radial directions, Fig. S2c. The bacterial concentration peaks at $r \sim 0\,\mu\text{m}$, and decays quickly with the increase in $r$, Fig. S2a. The absolute average value of the radial component of velocity, obtained by taking the population average over 90 sec, is $|v_r| = 4\,\mu\text{m/s}$. The azimuthal component is zero, $v_\phi = 0$. When the initial concentration is raised to $c_v \approx 0.16 \times 10^{14}\,\text{m}^{-3}$, $\Phi_v \approx 0.5 \times 10^{-4}$ ( $c_p \approx 0.85 \times 10^{14}\,\text{m}^{-3}$, $\Phi_p \approx 2.6 \times 10^{-4}$ ), the bacteria begin to accumulate in the center [3], Fig. S2d. With time, more bacteria are attracted towards the core, forming an immobilized dense cluster, as shown for $t = 50$ s in Fig. 2e and $t = 70$ s in Fig. S2f; here the time $t = 0$ s is defined as the starting point of tracking, which was 60 sec after cooling the sample down from the isotropic to the nematic phase, to ensure that all the samples are in a homogeneous deep nematic state.

**Circular vortex.** For $c_v \approx 0.8 \times 10^{13}\,\text{m}^{-3}$, $\Phi_v \approx 2.4 \times 10^{-5}$, the circular pattern with a pure bend, $\phi_0 = 90°$, guides the bacteria along circular trajectories of different radii, Fig. S2b,2r. The average absolute value of the azimuthal velocity is $|v_\phi| \approx 7.7\,\mu\text{m/s}$, but the number of bacteria swimming in the clockwise and counter-clockwise directions is approximately the same and there is no net flow along the azimuthal direction. The local bacterial concentration $c_p$ depends on the radial distance $r$ very differently from the radial case, Fig. S2b, since the bacteria do not approach the vortex core. The radius $r_{\text{empty}}$ of the empty central region increases with time, from $r_{\text{empty}} \approx 25\,\mu\text{m}$ at $t = 0\,\text{s}$ to $r_{\text{empty}} \approx 33\,\mu\text{m}$ at $t = 300\,\text{s}$, Fig. S2b, thus indicating a net outward radial flow that pushes the bacteria to the periphery of the circular vortex. Despite a high level of noise in data for the particular geometry described by Fig. S2b, the increase in $r_{\text{empty}}$ is statistically significant; in all 5 samples tested with a patterned circular vortex, $r_{\text{empty}}$ increases by more than $5\,\mu\text{m}$.

**Analytical solution for spiral vortices.** Transport of bacteria in LC is governed by two coupled advection-diffusion equations for the concentrations $c^\pm$ of bacteria swimming parallel to the director $\hat{\mathbf{n}}_0$ in clockwise ($c^+$) or counterclockwise ($c^-$) directions:

$$n_{0r} = \cos\phi_0, \tag{S1}$$

$$n_{0\phi} = \sin\phi_0. \tag{S2}$$

Here $\phi_0$ is the pre-described tilt angle with respect to the radial direction. We consider a steady state axisymmetric situation which does not depend on the polar angle $\phi$. Eq. (3) assume the form

$$\frac{1}{r}\partial_r r(-V_0 \hat{n}_r c^+ + v_r c^+) = -\frac{c^+ - c^-}{\tau} + D_c \frac{1}{r}\partial_r r \partial_r c^+ , \quad (S3)$$

$$\frac{1}{r}\partial_r r(V_0 \hat{n}_r c^- + v_r c^-) = -\frac{c^- - c^+}{\tau} + D_c \frac{1}{r}\partial_r r \partial_r c^- . \quad (S4)$$

As in the main text, here $c^\pm$ are the concentrations of bacteria swimming clock/counterclockwise, $\hat{n}$ is the actual director, $V_0$ is the bacterium swimming speed, $D_c$ is the diffusion coefficient, and $\tau$ is the direction reversal time (if a bacterium reverses a direction, it leaves the population $c^+$ and joins $c^-$ or vice versa).

In what follows, we consider the $\tau \to \infty$ limit (a very large reversal time). The equations for $c^\pm$ decouple. Let us assume that one of the populations swims away from the center (e.g. for the experimental configurations as in Fig.3 it is $c^-$) and consider only the population $c^+$. We can integrate stationary Eq. (S3) and obtain

$$-rV_0 n_r c^+ + r v_r c^+ = r D_c \partial_r c^+ + C_1 . \quad (S5)$$

The constant $C_1$ is zero in order to obey the condition at $r \to \infty$. In the steady state there is no radial flow, i.e. $v_r = 0$. Thus, Eq. (S5) simplifies to

$$-V_0 n_r c^+ = D_c \partial_r c^+ . \quad (S6)$$

For a dilute case, the experiment demonstrates that the actual director $\hat{n}$ coincides with the prescribed director $\hat{n}_0$ with $c^+ \sim \exp(-\frac{rV_0 \cos\phi_0}{D_c})$. This solution shows a maximum concentration at $r = 0$ and corresponds to the experiment with individual bacteria.

Now let us consider the collective effects. Since the bacteria exert dipolar stresses on the surrounding fluid, they produce an active stress $\boldsymbol{\sigma}_{act}$ in the form [4]

$$\boldsymbol{\sigma}_{act} = -\lambda \mathbf{Q} c^+ , \quad (S7)$$

where $\mathbf{Q}$ is the tensor order parameter of the nematic, $\mathbf{I}$ is the identity matrix, $\lambda > 0$ is the dipolar strength (sign "$-$" corresponds to pushers such as bacteria). The active stress will generate the realigning active force $\mathbf{F}_{act} = \nabla \cdot \boldsymbol{\sigma}_{act}$. To describe this effect and ensure $\hat{n}\hat{n} = 1$, we apply a relaxation equation in the Landau-Lifshitz-Gilbert form [5] for the director in the bulk,

$$\partial_t \hat{\mathbf{n}} = -\hat{\mathbf{n}} \times \hat{\mathbf{n}} \times \left( \gamma_s \hat{\mathbf{n}}_0 - 2a\mathbf{F}_{act} + K\Delta\hat{\mathbf{n}} \right). \tag{S8}$$

Eq. (S8) relaxes the actual director $\hat{\mathbf{n}}$ towards the direction defined by a combination of the preferred direction $\hat{\mathbf{n}}_0$ and active force and automatically fulfils the normalization condition $\hat{\mathbf{n}}\hat{\mathbf{n}} = 1$. Here $\gamma_s$ is the director relaxation rate dependent on the director anchoring strength at the bounding plates and $a$ is a coefficient dependent on the surface anchoring strength at the bacterial body. In Eq. (S8) we neglect for simplicity the elastic response of the nematic and set the elastic constant $K = 0$. A stationary solution to Eq. (S8) for $K = 0$ is of the form

$$\hat{\mathbf{n}} = \frac{\hat{\mathbf{n}}_0 - 2a\mathbf{F}_{act}/\gamma_s}{\left|\hat{\mathbf{n}}_0 - 2a\mathbf{F}_{act}/\gamma_s\right|}. \tag{S9}$$

Note that due to director identity, $\mathbf{n} \leftrightarrow -\mathbf{n}$, the definitions with opposite signs are both valid, as in Ref. [6]. Since in 2D the Q-tensor is of the form

$$\mathbf{Q} = \begin{pmatrix} \cos(2\phi) & \sin(2\phi) \\ \sin(2\phi) & -\cos(2\phi) \end{pmatrix} \tag{S10}$$

and the radial component of the tensorial divergence is of the form $r^{-2}\partial_r r^2$, we obtain for $n_r$:

$$n_r = \cos(\phi) = \frac{\cos(\phi_0) + \eta\cos(2\phi)r^{-2}\partial_r r^2 c^+}{\sqrt{(\cos(\phi_0) + \eta\cos(2\phi)r^{-2}\partial_r r^2 c^+)^2 + (\sin(\phi_0) + \eta\sin(2\phi)r^{-2}\partial_r r^2 c^+)^2}}, \tag{S11}$$

where $\eta = 2a\lambda/\gamma_s$. By substituting Eq. (S11) into Eq. (S6) we obtain an implicit first-order nonlinear ordinary differential equation for $c^+$ which can be solved numerically. The solutions can be quantified by the total amount of bacteria $M = 2\pi\int_0^\infty rc^+ dr$ (or initial condition for $c^+$ at $r = 0$). Plot of $c^+$ vs $r$ for different values of total amount of bacteria is shown Fig. S6a. Respectively, the position of the maximum moves right with the increase in $\phi_0$, see Fig. S6b. The obtained results exhibit qualitative agreement with experiment and more detailed agent-based and advection-diffusion models of living liquid crystal.

**Numerical solution of the Landau-Lifshitz-Gilbert equation for agent-based simulation.** The evolution of pre-patterned director is described by the Landau-Lifshitz-Gilbert equation in the form Eq. (S8). However, its numerical integration is computationally challenging because Eq. (S8) is higly nonlinear and not too well behaved in topological defects. Instead of Eq. (S8) we solved well-behaved regularized Ginzburg-Landau equation for the complex order parameter $\psi = n_x + in_y$.

$$\partial_t \psi = b(\psi - i|\psi|^2 \psi) + K\nabla^2 \psi + i\psi \, \text{Im}(\psi^*(\gamma_B \psi_0 - 2a\chi_{act})) \,. \tag{S12}$$

Here the first term with $b > 0$ ensures $|\psi|^2 = 1$ except the location of topological defects where $\psi = 0$. The last term implemets the Landau-Lifshitz-Gilbert dynamics. Here $\psi_0 = n_{0x} + in_{0y}$ and $\chi_{act} = F_{act,x} + iF_{act,y}$. Equation (S12) was solved in a rectangular domain by qausi-spectral split-step method based on the fast Fourier transformation, see Refs. [7,8]. In order to find the value of $\psi$ between the computational grid points, qubic spline interpolation was used. The algorithm is implemented in MATLAB computational package.

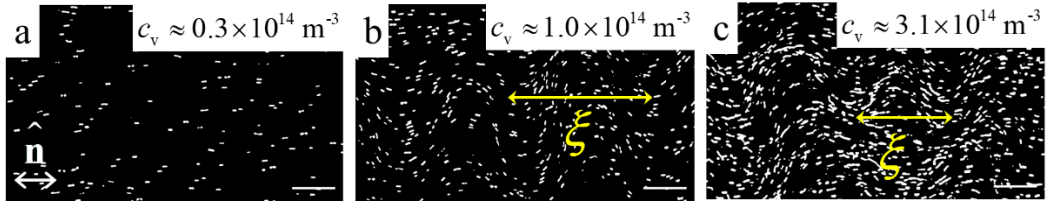

**Figure S1.** Optical microscopy textures demonstrating effects of bacterial concentration on their collective dynamics. (a) In dilute dispersions, $c_v \approx 0.3 \times 10^{14}$ m$^{-3}$, $\Phi_v \approx 0.9 \times 10^{-4}$, bacteria swim parallel to the imposed director shown by the double-headed arrow; (b) At $c_v \approx 1.0 \times 10^{14}$ m$^{-3}$, $\Phi_v \approx 3 \times 10^{-4}$, bacterial trajectories start to undulate; (c) At high concentrations, $c_v \approx 3.1 \times 10^{14}$ m$^{-3}$, $\Phi_v \approx 9.3 \times 10^{-4}$, period $\xi$ of undulations shortens . Scale bar $50\,\mu\text{m}$.

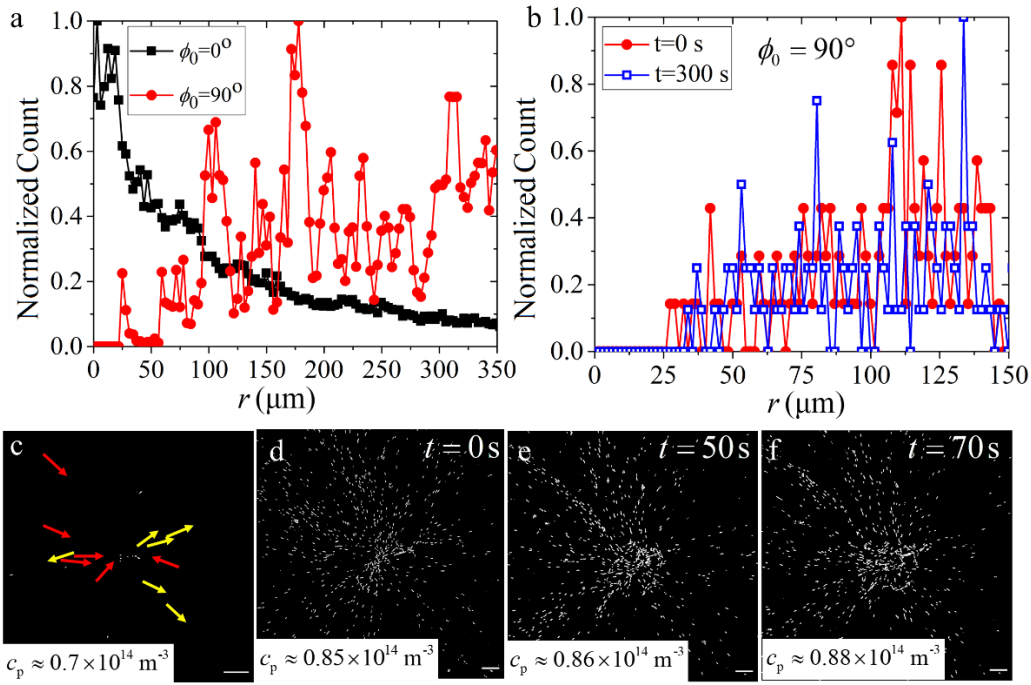

**Figure S2.** Distribution and speed of bacteria in pure splay ($\phi_0 = 0$, aster) or pure bend ($\phi_0 = 90°$, center) patterns. (a) Distribution of bacteria along the radial direction for $\phi_0 = 0$ and $\phi_0 = 90°$. The analysis was done by dividing the area of interest into multiple narrow annuli of width $\Delta r = 3\,\mu m$ and counting the number of bacteria in each annulus averaged throughout the time interval [0, 90 s]. The data are then normalized by dividing the number of bacteria in each annulus by the number of bacteria in the most populated annulus. (b) Normalized distribution of bacteria along the radial direction for $\phi_0 = 90°$, tracked within a time interval [0, 300 s]. (c) Optical microscopy texture of bacteria swimming in and out of a pure splay pattern ($\phi_0 = 0$) for $c_p \approx 0.7 \times 10^{14}\,m^{-3}$ ($c_v \approx 0.8 \times 10^{13}\,m^{-3}$, $\Phi_v \approx 2.4 \times 10^{-5}$). The red vectors show bacteria swimming inward and the yellow vectors show bacteria swimming outward. (d)-(f) Time evolution of bacterial condensation at the center of a radial splay pattern ($\phi_0 = 0$). Scale bar $50\,\mu m$.

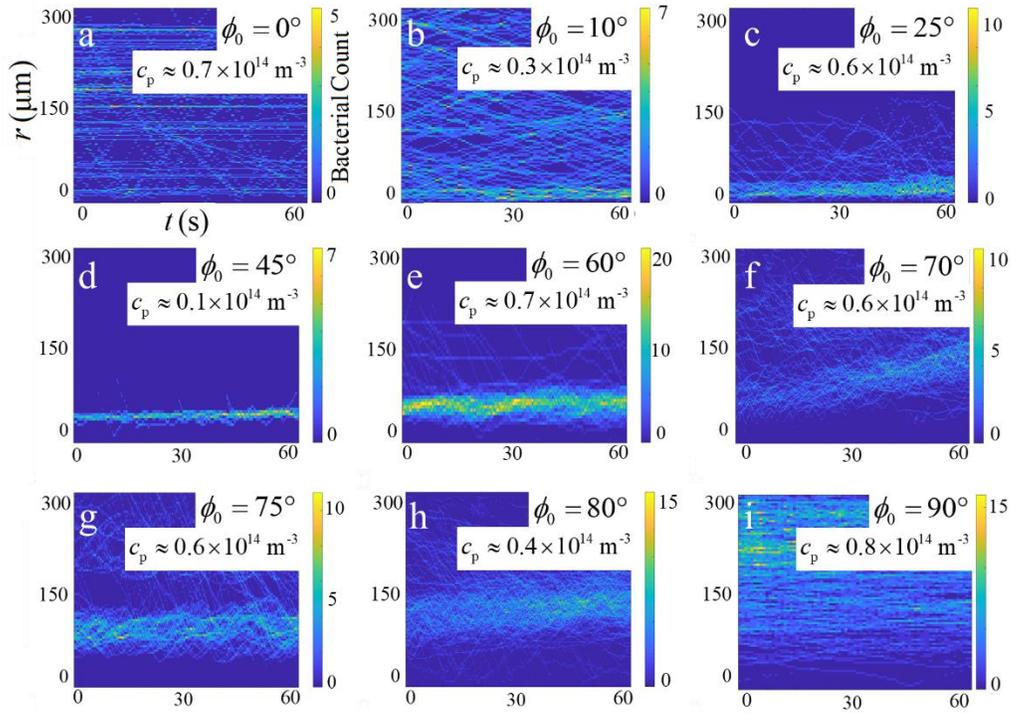

**Figure S3.** Kymograph of bacteria along the radial direction tracked within the time interval [0, 60 s] for (a) $\phi_0 = 0$, (b) $\phi_0 = 10°$, (c) $\phi_0 = 25°$, (e) $\phi_0 = 45°$, (e) $\phi_0 = 60°$, (f) $\phi_0 = 70°$, (g) $\phi_0 = 75°$, (h) $\phi_0 = 80°$ and (i) $\phi_0 = 90°$. All samples are prepared with the same initial concentration $c_V \approx 0.8 \times 10^{13}$ m$^{-3}$, $\Phi_V \approx 2.4 \times 10^{-5}$. The analysis was done by dividing the area of interest into multiple narrow annuli of width $\Delta r = 3\,\mu$m and counting the number of bacteria within each annulus using particle tracking. The color bar shows the number of bacteria detected within each annulus of width $\Delta r = 3\,\mu$m. The kymographs demonstrate that the spiral vortices with spiral angle $0 < \phi_0 < 90°$ serve as condensors of bacteria into toroidal groups, concentrating them within relatively narrow toroidal bands that rotate unidirectionally.

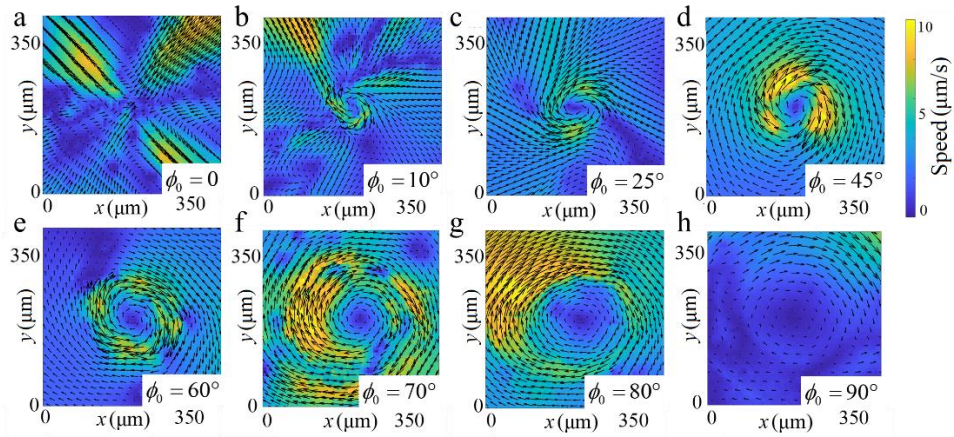

**Figure S4.** Velocity map of bacteria averaged over a time interval [0, 60s] and moving in (a) $\phi_0 = 0$, (b) $\phi_0 = 10°$, (c) $\phi_0 = 25°$, (e) $\phi_0 = 45°$, (e) $\phi_0 = 60°$, (f) $\phi_0 = 70°$, (g) $\phi_0 = 80°$ and (h) $\phi_0 = 90°$ patterned director fields obtained from PIV. The color bar shows the speed $v = \sqrt{v_r^2 + v_\phi^2}$ μm/s. All samples are prepared with the same initial concentration $c_v \approx 0.8 \times 10^{13}$ m$^{-3}$, $\Phi_v \approx 2.4 \times 10^{-5}$.

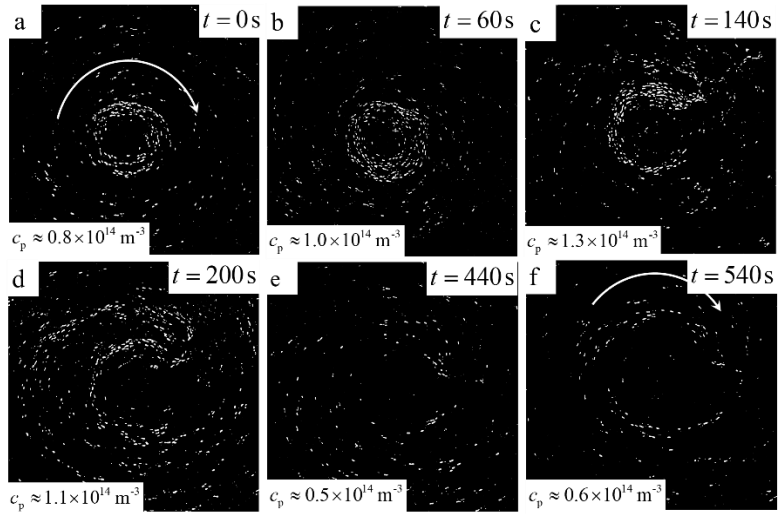

**Figure S5.** Optical microscopy textures of time sequence of undulations and population change in $\phi_0 = 75°$ pattern. (a). Circular swirling. (b) Circular swirling with an increased local concentration of bacteria; start of weak undulations. (c) Rupture of the swirl. (d,e) Escape of some bacteria. (f).

Restoration of a circular swirl with a radius larger than that of the original swirl in part (a). Scale bar $50\,\mu m$.

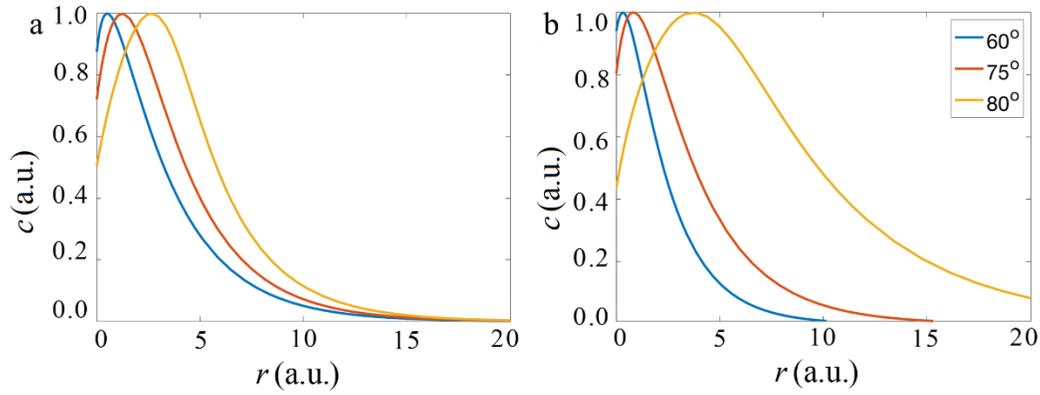

**Figure S6.** Plots of normalized concentration $c = c^- / c_{max}$ vs radial distance $r$ for (a) increasing bacterial concentrations (yellow is max bacterial concentration) for $\phi_0 = 70°$ and (b) increasing prescribed tilt angle $\phi_0$, where $\phi_0 = 60°$ (blue), $\phi_0 = 70°$ (red), and $\phi_0 = 80°$ (yellow). For both plots, $D_c = 1\,\mu m^2/s$, $\eta = 1$, $V_0 = 1\,\mu m/s$.

## SI References

**Supplementary Movies**

**Movie S1.** Movie of dilute dispersion ($c_p \approx 0.5 \times 10^{12}$ m$^{-3}$ $< c_p^{\min}$, $c_v \approx 0.8 \times 10^{12}$ m$^{-3}$, $\Phi_v \approx 2.4 \times 10^{-6}$) of bacteria following the director field for $\phi_0 = 25°$. Scale bar $50$ μm. The video is recorded at 1 frame/s and played back at 10 frames/s.

**Movie S2.** Movie of bacteria swimming in intermediate spiral-circular state ($c_p = c_p^{\min} \approx 0.2 \times 10^{13}$ m$^{-3}$, $c_v \approx 0.8 \times 10^{12}$ m$^{-3}$, $\Phi_v \approx 2.4 \times 10^{-6}$) for $\phi_0 = 25°$. Scale bar $50$ μm. The video is recorded at 1frame/s and played back at 10 frames/s.

**Movie S3.** Movie of bacteria swimming in a circular swirl ($c_p \approx 0.2 \times 10^{14}$ m$^{-3}$ $> c_p^{\min}$, $c_v \approx 0.8 \times 10^{13}$ m$^{-3}$, $\Phi_v \approx 2.4 \times 10^{-5}$) for $\phi_0 = 25°$. Scale bar $50$ μm. The video is recorded at 1 frame/s and played back at 10 frames/s.

**Movie S4.** Movie of unidirectional circular swirling of bacteria for $\phi_0 = 75°$ where $c_p \approx 0.6 \times 10^{14}$ m$^{-3}$ ($c_v \approx 0.8 \times 10^{13}$ m$^{-3}$, $\Phi_v \approx 2.4 \times 10^{-5}$). Scale bar $50$ μm. The video is recorded at 1 frame/s and played back at 10 frames/s.

**Movie S5.** Numerical simulation for $\phi_0 = 45°$ with different number of bacteria. (a) $n_B$ = 4, 9, 16; (b) $n_B$ = 25, 36, 64; (c) $n_B$ = 100, 196, 256.

**Movie S6.** Numerical simulation of $n_B = 256$ bacteria in vortices with different $\phi_0$. (a) $\phi_0 = 0, 10°, 30°$; (b) $\phi_0 = 45°, 50°, 60°$; (c) $\phi_0 = 70°, 80°, 90°$.

**Movie S7.** Numerical simulation of the advection-diffusion model in the regime of undulation instability for $\phi_0 = 85°$. (a) Director orientation (lines) and amplitude (color); (b) Bacterial concentration difference $w = c^+ - c^-$; (c) Fluid velocity; (d) Total bacterial concentration $c = c^+ + c^-$.